\documentclass[11pt]{article}
\usepackage[a4paper,bottom=4.2cm,top=1cm,head=3cm,width=18cm,dvipdfm]{geometry}
\evensidemargin=\oddsidemargin

\usepackage[dotinlabels]{titletoc}
\usepackage{titlesec}
\usepackage[numbers,sort&compress]{natbib} 

\usepackage{xcolor}
\usepackage{mathrsfs}
\usepackage{amsmath}
\usepackage{amssymb}
\usepackage{bm}
\usepackage{amsfonts}
\usepackage{subfigure}
\usepackage{feynmp}
\usepackage{extarrows}
\usepackage{slashed}
\usepackage{graphicx}
\usepackage{dcolumn}
\usepackage{verbatim}
\usepackage{here}
\usepackage{multirow} 
\allowdisplaybreaks[4]
\usepackage{indentfirst}
\usepackage{isodateo}
\usepackage{enumerate}
\usepackage[low-sup]{subdepth}

\numberwithin{equation}{section}
\renewcommand{\thefootnote}{\arabic{footnote}}
\usepackage{isodateo}
\usepackage[CJKbookmarks=true, bookmarksnumbered=true,bookmarksopen=true,]{hyperref}
\hypersetup{colorlinks,%
	linkcolor=blue,
	citecolor=blue,
urlcolor=blue}

\graphicspath{{figs/}}

\newcommand{\fr}[2]{\mbox{$\frac{\,{#1}\,}{#2}$}}

\newcommand{\be}{\begin{equation}}
\newcommand{\ee}{\end{equation}}
\newcommand{\bea}{\begin{eqnarray}}
\newcommand{\eea}{\end{eqnarray}}

\def\ede{\end{equation}}
\def\bga{\begin{aligned}}
\def\eda{\end{aligned}}
\newcommand{\beq}{\begin{equation}}
\newcommand{\eeq}{\end{equation}}
\newcommand{\bq}{\begin{equation}}
\newcommand{\eq}{\end{equation}}
\newcommand{\ba}{\begin{array}}
\newcommand{\ea}{\end{array}}
\newcommand{\beqa}{\begin{eqnarray}}
\newcommand{\eeqa}{\end{eqnarray}}
\newcommand{\beqs}{\begin{subequations}}
\newcommand{\eeqs}{\end{subequations}}

\def\nn{\nonumber}

\def\({\left(}
\def\){\right)}

\def\leqq{\leqslant}

\def\End{\end{document}}
\def\d{\text{d}}
\def\ii{{\tt i}}
\def\over{\overline}

\def\be{\beta}

\def\mX{m_X^{}}
\def\hs{\hspace*{0.3mm}}
\def\hsx{\hspace*{0.5mm}}
\def\hsm{\hspace*{-0.3mm}}
\def\hsmx{\hspace*{-0.5mm}}

\def\End{\end{document}}

\begin{document}

 \thispagestyle{empty}
 \renewcommand{\thefootnote}{\fnsymbol{footnote}}
 \setcounter{footnote}{0}
 \titlelabel{\thetitle.\quad \hspace{-0.8em}}
\titlecontents{section}
              [1.5em]
              {\vspace{4mm} \large \bf}
              {\contentslabel{1em}}
              {\hspace*{-1em}}
              {\titlerule*[.5pc]{.}\contentspage}
\titlecontents{subsection}
              [3.5em]
              {\vspace{2mm}}
              {\contentslabel{1.8em}}
              {\hspace*{.3em}}
              {\titlerule*[.5pc]{.}\contentspage}
\titlecontents{subsubsection}
              [5.5em]
              {\vspace{2mm}}
              {\contentslabel{2.5em}}
              {\hspace*{.3em}}
              {\titlerule*[.5pc]{.}\contentspage}
\titlecontents{appendix}
              [1.5em]
              {\vspace{4mm} \large \bf}
              {\contentslabel{1em}}
              {\hspace*{-1em}}
              {\titlerule*[.5pc]{.}\contentspage}


\vspace*{8mm}

\begin{center}
	{\Large\bf Probing Light Inelastic Dark Matter from Direct Detection}
	
\vspace*{8mm}
	
{\large\sc Hong-Jian He},$^{a,b}$\footnote{Email: hjhe@sjtu.edu.cn}~
{\large\sc Yu-Chen Wang},$^{a}$\footnote{Email: wangyc21@sjtu.edu.cn}~
{\large\sc Jiaming Zheng},$^{a}$\footnote{Email: zhengjm3@gmail.com}
	
\vspace*{4mm}
	
$^a$\,T.\ D.\ Lee Institute $\&$ School of Physics and Astronomy, \\
Key Laboratory for Particle Astrophysics and Cosmology,\\
Shanghai Key Laboratory for Particle Physics and Cosmology,\\
Shanghai Jiao Tong University, Shanghai, China
\\[1mm]
$^b$\,Physics Department and Institute of Modern Physics,
Tsinghua University, Beijing, China;
\\ 
Center for High Energy Physics, Peking University, Beijing, China

\vspace*{20mm}
\end{center}

\vspace*{3mm}

\begin{abstract}
\baselineskip 17pt   
\noindent
For dark matter (DM) direct detections, 
the kinematic effects such as those of the inelastic scattering 
can play important role in light DM searches.\ 
The light DM detection is generally difficult because of its small recoil energy.\  
But the recoil energy of the exothermic inelastic DM scattering could 
exceed the detection threshold due to the contribution from the DM mass-splitting,
making the direct detection of sub-GeV DM feasible.\ 
In this work, we systematically study signatures of the light exothermic inelastic DM 
from the recoil spectra including both the DM-electron scattering and Migdal effect.\ 
Such inelastic DM has mass around (sub-)GeV scale with 
DM mass-splitting of $O(1\!-\!10^2)$keV.\ 
We analyze the direct detection sensitivities to such light inelastic DM.\ 
For different inelastic DM masses and mass-splittings, we find that
the DM-electron recoil and Migdal effect can contribute significantly 
and differently to the direct detection signatures.\ 
The DM-lepton and/or DM-quark interactions may vary for different DM models, 
and their interplay leads to a diversity in the recoil spectra.\ 
Hence, it is important to perform a {\it combined analysis} to include both 
the DM-electron recoil and Migdal effect.\ 
We further demonstrate that this analysis has strong impacts 
on the cosmological and laboratory bounds for the inelastic DM. 
\\[5mm]
Physics of the Dark Universe 46 (2024) 101670 [\,arXiv:2403.03128\,].
\end{abstract}

  \newpage
\renewcommand{\thefootnote}{\arabic{footnote}}
\setcounter{footnote}{0}
\setcounter{page}{2}

\tableofcontents

\setcounter{footnote}{0}
\renewcommand{\thefootnote}{\arabic{footnote}}

\baselineskip 18pt

\vspace*{10mm}
\section{Introduction}
\vspace*{1.5mm}
\label{sec:intro}
\label{sec:1}

The origin and nature of dark matter (DM) have remained largely unknown so far
after decades of dedicated efforts through the direct, indirect and
collider searches.\ Such detection methods depend on how the DM interacts with
the standard model (SM) particles.\ 
The DM models have provided various DM-SM interactions and are under close 
inspection via different direct search strategies.\ 
The searches for nuclear recoil (NR) have set stringent limits 
on the DM-nucleon scattering cross 
sections\,\cite{SuperCDMS:2017mbc}-\cite{LZ:2022ufs}
on the GeV-TeV scale DM.\ 
For lighter dark matter, the nuclear recoil energy falls below their detection thresholds.\ 
In contrast, the light DM scattering with electron can produce significant recoil energy for 
the electron\,\cite{SuperCDMS:2020ymb}-\cite{SENSEI:2023zdf}.\
The electron recoil (ER) signal is thus an important target for searching
the (sub-)GeV DM candidates\,\cite{Essig:2011nj}.\  
The electron recoil signal can also probe DM-nucleon interaction through 
the Migdal effect\,\cite{Migdal}\cite{Ibe:2017yqa}.\ 
When the DM scatters with an atomic nucleus, 
the nucleus gains a velocity relative to the electron cloud 
and may induce atomic ionization and excitation.\ 
For sub-GeV DM, this process can generate keV-scale electron recoil signals\,\cite{Dolan:2017xbu}-\cite{Wang:2021oha}
that are detectable in the direct detection 
experiments\,\cite{CDEX:2019hzn}-\cite{SuperCDMS:2023sql}.\

A main difficulty in detecting light DM is that 
the kinetic energy of the incoming DM particle is insufficient to trigger detectable signals.\  
Potential kinematic effects, such as those induced 
by exothermic inelastic dark matter\,\cite{ineDMold}-\cite{Bramante:2020zos}
may avoid this difficulty.\ 
In such a scenario, the DM particles consist of more than one mass eigenstate.\ 
When a heavier DM state scatters with a target atom, it converts to 
the lighter state and deposits energy from the mass-splitting 
into the recoil energy of the target.\ 
As long as the recoil energy induced by the mass-splitting 
is above the detection threshold (usually at keV scale), 
there is no need for a large DM kinetic energy, and thus 
the DM particles can be rather light.\ 
Moreover, this deposited energy can lead to significant changes in the recoil spectra 
of direct detection experiments.\ 
For DM-electron scattering, it results in a peak-like ER spectrum\,\cite{He:2020wjs}\cite{He:2020sat}.\ 
This gives a highly distinctive and detectable signal 
for DM direct detection experiments.\ 
For DM-nucleon scattering, the deposited energy enhances the 
Migdal effect\,\cite{Bell:2021zkr}, 
especially for sub-GeV DM whose kinetic energy alone 
is insufficient to excite the atomic electrons.\ 
This leads to a higher experimental sensitivity than the conventional elastic DM case.\
The experimental collaborations for DM direct detections recently published 
the electron recoil data with low threshold and 
high sensitivity\,\cite{PandaX-II:2021nsg}-\cite{XENON:2022ltv}\cite{LZ:2023poo}.\ 
Hence, it is ideal to use these datasets and analyze such kinematic effects 
of DM interactions.\

\vspace*{0.8mm}

Conventionally, the DM-nucleon and DM-electron interactions are constrained independently. For DM-nucleon spin-independent (SI) interactions, constraints are usually derived by assuming identical contributions from neutron and proton.\ This enhances the DM-nucleus scattering cross section by a coherence factor of $A^2$, where $A$ is the number of nucleons.\
However, these assumptions do not always hold for different DM models.\ 
First, the DM particle in general may interact with both nucleons and electrons.\ 
Second, the DM-neutron and DM-proton interactions may not be identical or 
coherent within the nucleus, so they may induce a coherence factor other than $A^2$,
which will change not only the overall DM-nucleon event rates,
but also the relative strength between DM-electron and DM-nucleon interactions.\ 
These may significantly change the constraints on different DM models.\ 
Thus, although the DM-nucleon and DM-electron interactions have already been separately studied in literature, a joint analysis including both effects should be 
necessary and important.

\vspace*{0.8mm}

In this work, we study direct detection signatures of the light 
exothermic inelastic DM with keV-scale mass-splitting.\ 
Since electrons, neutrons, and protons may couple differently to the dark sector,
we consider various combinations of typical couplings in the present analysis.\ 
For each case, we compute the inelastic DM-electron and DM-nucleus scattering rates, including the induced Migdal effect.\
We demonstrate that the DM-electron scattering spectrum is highly dependent 
on the mass-splitting, whereas the
Migdal effect induced by DM-nucleus scattering 
is enhanced by the energy injection 
from the de-excitation of the heavier DM state.\
We further derive the bounds for each case by confronting
the XENONnT ER data\,\cite{XENON:2022ltv} 
and XENON1T S2-only data\,\cite{XENON:2019gfn} 
with the joint effects of DM-electron and DM-nucleon interactions.\ 
We demonstrate that the inelastic DM-electron scattering 
creates a unique peak-like signal in the electron recoil spectrum,
and in DM-nucleus scatterings the exothermic effect largely enhances 
the signal from Migdal effect.\  
Including both effects, we can obtain strong bounds on the (sub-)GeV inelastic DM.\  
We also show that these direct detection bounds 
can be readily applied to concrete models of inelastic DM.

\vspace*{0.8mm}

This paper is organized as follows.\
In Section\,\ref{sec:2}, we present a generic effective theory formulation of  
the light exothermic inelastic DM interactions.\ 
We consider both the DM-electron scattering and 
DM-nucleus scattering with Migdal effect.\ For this we will present the 
benchmark scenarios of the dark charge assignments.\  
In Section\,\ref{sec:3}, we study the electron recoil signatures 
from both the direct DM-electron scattering and the Migdal effect (induced by
DM-nucleus scattering).
In Section\,\ref{sec:4}, we demonstrate how direct detection experiments 
(such as XENON1T and XENONnT) can constrain the parameter space
of the light inelastic DM.\ 
We present an explicit model realization of the light inelastic DM 
in Section\,\ref{sec:5}, and analyze constraints from both 
the cosmological and laboratory measurements.\ 
Finally, we conclude in Section\,\ref{sec:6}.

\vspace*{2mm}
\section{Inelastic Dark Matter Interactions for Exothermic Scattering}
\label{sec:DMmodels}
\label{sec:2}
\vspace*{1mm}

In this section, we present a generic effective theory formulation of  
the light inelastic DM interactions.\ With this, we will further study the interplay 
between the DM-nucleon and DM-electron interactions.\

We consider the exothermic inelastic scattering process, 
\begin{eqnarray}
	X' + f \rightarrow X + f \hs,
\end{eqnarray}
where $f$ is a SM fermion, such as an electron or a quark.\ 
The $X'$ and $X$ denote the heavier and lighter component of a scalar DM, respectively.\ 
In such a process, the energy release is the mass-splitting between the two components
of DM, $\Delta m \equiv m_{X'}-m_X$.\ 
For the fermionic DM case, we denote the lighter (heavier) component as $\chi~(\chi')\hs$.

Here $X$ and $X'$ are real scalar fields and form a complex scalar 
$\,\hat{X}\!\equiv\!(X\!+\ii X')/\hsm\sqrt{2\,}\,$ 
in the limit $\Delta m\!\to\! 0\hs$.\ 
Similarly, $\chi$ and $\chi'$ are Weyl spinors and can form a pseudo-Dirac spinor $\hat{\chi}\equiv (\chi_1^{}, \chi_2^\dagger)^T$, where 
$\chi_{1,2}^{}\hsm \equiv(\chi\mp\chi')/\hsm\sqrt{2\,}\hs$.\ 
When the situation is not sensitive to the spin-statistics of the DM, 
we will only use the notation $(X,\hs X')$ 
for both the scalar and fermionic DM.\ 
Consider that the DM particles interact with the visible sector through 
a vector mediator whose effect can be described by the generic dimension-6 
effective operators at low energy,
\begin{equation}\label{eq:dim6}
{\cal O}\,=\, \frac{c}{\,M^2\,} J_{X\mu}^{} J_{\rm vis}^\mu\,, 
\end{equation}
where the mediator mass $M$ serves as the cutoff scale
and the operator ${\cal O}$ is invariant under the SM gauge group.\ 
In the above, we denote the dark current $J_{X\mu}^{}$ for scalar DM $(X,\hs X')$, 
and this notation is replaced by $J_{\chi\mu}^{}$ for the fermionic DM 
$(\chi,\hs \chi')$.\ 
The two dark currents $J_{X}^{\hs\mu}$ and $J_{\chi}^{\hs\mu}$ 
take the following forms:
\\[-11mm]
\beqs 
\begin{align}
J_{X}^{\hs\mu}
&=\,\ii(\hat{X}^*\partial^\mu\hat{X}\!-\!\hat{X}\partial^\mu\hat{X}^*)
=(X'\partial^\mu X\!-\!X'\partial^\mu X) \hs ,	
\\
J_{\chi}^{\hs\mu} 
&=\,\overline{\hat{\chi}}\gamma_\mu\hat{\chi}
= \ii(\chi'^\dagger\bar\sigma_\mu \chi\!-\!\chi^\dagger\bar\sigma_\mu \chi') \hs .
\end{align}
\eeqs 
The mass-splitting between the dark components 
can be induced by the spontaneous symmetry-breaking mechanism 
that generates the mass of the vector mediator.\ 
This will be shown by a concrete UV completion model in Sec.\,\ref{sec:models}.\
The current for the visible sector can be separated into leptonic and hadronic parts, $J_{\rm vis}^\mu\!=\!J_{\ell}^\mu\!+\!J_q^\mu\hs$, and takes the general form:
\begin{subequations}
\label{eq:J-lep+q}
\begin{align}
\label{eq:J-lep}
J_{\ell}^\mu &=\, 
c_L^{}\bar{L}\gamma^\mu L + c_e^{}\bar{e}_R^{}\gamma^\mu e_R^{}\,, 
\\
\label{eq:J-q}
J_{q}^\mu &=\, 
c_Q^{}\bar{Q}\gamma^\mu Q + c_u^{}\bar{u}_R^{}\gamma^\mu u_R^{} 
+ c_d^{}\bar{d}_R^{}\gamma^\mu d_R^{}\,,
\end{align}
\end{subequations}
where $(c_L^{},\,c_e^{})$ and $(c_Q^{},\,c_u^{},\,c_d^{})$ denote respectively 
the lepton and quark couplings to the vector mediator.\ 
For the direct detection analysis, 
we express the general currents \eqref{eq:J-lep+q} 
for electron and $(u,\hs d)$-quarks as follows:
\\[-11mm]
\begin{subequations}
\begin{align}
J_{e}^\mu &=\, c^V_e\bar{e}\gamma^\mu e + c^A_e\bar{e}\gamma^\mu\gamma^5 e\,,
\\
J_{q}^\mu &=\, c^V_q\bar{q}\gamma^\mu q + c^A_q\bar{q}\gamma^\mu\gamma^5 q\,,
\end{align}
\end{subequations}
where the quark field $\,q\!=\!u,d\hs$.\ 
Hereafter, we will only consider the vector current of quarks 
in the DM-nucleon scattering for the following reasons.\ 
The nuclear expectation value of the quark axial current is proportional to the spin fractions in the target nucleus\,\cite{Engel:1989ix} and is much smaller than the nuclear expectation value of the quark vector current in general cases because the latter is enhanced by the number of nucleons.\    
Besides, the DM current in the effective operator is chosen to be a vector current,
and its couplings with quark (electron) axial currents only contribute to the DM-nucleus (electron) scattering amplitudes with terms suppressed by the small DM-velocity.\ Hence, we will work at the leading order of the small momentum transfer and neglect the quark or electron axial current in the direct detection analysis.\
The DM-quark interaction matches to the spin-independent DM-nucleon interactions\,\cite{Jungman:1995df}:
\begin{equation}
\frac{c}{\,M^2\,} J_{X\mu}^{}
( q_p^{}\hs\bar{p}\gamma^\mu p + q_n^{} \bar{n}\gamma^\mu n )\,,
\end{equation}
where the coupling coefficients (dark charges) for protons and neutrons are
$\,q_p^{}\!=\! 2c^V_u\hsm\!+\hsm c^V_d\hs$ and  
$\,q_n^{}\!=\!c^V_u\!+\hsm 2c^V_d$.\ 
In the limit of $\Delta m\!\to 0\,$,
the DM-nucleon scattering cross section 
for $X'\!+\hsm (n,p)\!\to\! X\!+\hsm (n,p)\,$ can be expressed as
\begin{equation}
\label{eq:sigma-X'np-Xnp}
\sigma_{n,p}^{} \,\simeq \frac{\,|c|^2 q_{n,p}^2 \mu_{n,p}^2\,}{\pi M^4}
= \frac{\,4\alpha q_{n,p}^2 \mu_{n,p}^2\,}{\Lambda^4}\,,
\end{equation}
where $\hs\mu_{n,p}^{}\!=\hsm m_{n,p}^{}m_X/(m_{n,p}^{}\!+\!m_X^{})\hs$ 
is the reduced mass.\ 
In the second step of Eq.\eqref{eq:sigma-X'np-Xnp}, 
we have defined an effective cutoff scale 
$\Lambda\hsm\equiv\hsm M\hs (e/|c|)^{\frac{1}{2}}_{}$, where 
$\hs\alpha\!=\!e^2/4\pi\,$ denotes the fine structure constant and  
$e$ is the unit of electric charge.\ 
Then, the DM-nucleus ($N$) spin-independent scattering cross section takes the form:
\begin{equation}
\label{eq:sigma-N}
\sigma_{N}^{} \,\simeq\hs \frac{\,4\hs\alpha\hs q_N^2 \mu_{N}^2\,}{\Lambda^4}\,,
\end{equation}
where the DM-nucleus effective coupling
$q_N^{}(A,Z)\!=\!Z\hs q_p^{}+(A\hsm -\hsm Z)q_n^{}$ \cite{Jungman:1995df}\cite{Griest:1988ma}\cite{Lewin:1995rx}, 
and the reduced mass 
$\mu_{N}^{}\!=\!m_{N}^{}m_X^{}/(m_{N}^{}\hsm +\hsm m_X^{})$.\ 
The quantities $A$ and $Z$ denote the nucleon number 
and the electric charge of the nucleus, whereas $m_N^{}$ denotes the nucleus mass.\ 
We can ignore the nuclear form factor because for the parameter space considered 
in the present study, the momentum transfer in a DM-nucleus scattering 
is always small as compared to the inverse of the nuclear radius.\ 
We can also ignore the nuclear excitations as the energy transfer 
is relatively small.\ 
In parallel, the DM-electron scattering cross section is given by
\begin{equation}
\label{eq:sigma-e}
\sigma_e^{}\, \simeq\hs \frac{\,4\hs\alpha\hs q_e^2 \mu_e^2\,}{\Lambda^4}\,,
\end{equation}
where the dark charge of electron is $\,q_e^{}\!=c^V_e$.\ 
For the case $\,m_X^{}\!\gg m_e\hs$, the reduced mass is 
$\hs\mu_e\!\simeq m_e^{}\hs$.
Taking the ratio of cross sections of Eqs.\eqref{eq:sigma-N}-\eqref{eq:sigma-e},
we have
\beq 
\label{eq:ratio-Sigma-eN}
\frac{\sigma_e^{}}{\,\sigma_N^{}\,} \,\simeq\,
\(\!\frac{\,q_e^{}\hs\mu_e^{}\,}{\,q_N^{}\mu_N^{}\,}\!\)^{\!\!2}. 
\eeq 
Eq.\eqref{eq:ratio-Sigma-eN} shows 
that the interplay between DM-electron scattering and (spin-independent) DM-nucleus scattering can be encoded into the charge ratio between $q_e^{}$ and $q_{n,p}^{}\hsx$.\footnote{%
Note that Eq.\eqref{eq:ratio-Sigma-eN} alone does not determine the full ratio 
between their event numbers; other factors such as the electron transition 
probabilities will be discussed in Section.\,\ref{sec:3}.}\
Although this work is motivated 
by studying a dark sector with a vector mediator, the parameterization of 
the signal cross sections $(\sigma_{e}^{},\sigma_{n}^{},\sigma_{p}^{})$ 
applies to general spin-independent interactions.\ 
Hence, the analysis of the electron recoil spectrum 
in the following sections can be applied to other models with modified matching conditions between quark-level and nucleon-level interactions.\ 
In the following, we will consider several benchmark combinations of 
$\hs q_{e,n,p}^{}\!=\!0,\pm 1$ as shown in Table\,\ref{tab:benchmarks}.\ 
Note that the sign of $q_{e}^{}$ does not affect the signals.\ 
In the same table, we also indicate the corresponding
colored curve for each scenario as to be shown in the figures presenting  
the electron recoil spectra of Section\,\ref{sec:4}.\ 
The benchmark scenarios\,(a)-(d) have $\,q_e^{}\!=\!0\hs$, 
so the electron recoil signal arises only from the Migdal effect of DM-nucleus scattering.\ The benchmark\,(e) has $q_{n,p}^{}\!=\!0$ and its signal comes 
exclusively from the DM-electron scattering.\ 
For benchmarks\,(f)-(j), the signals arise from both DM-nucleus and DM-electron scattering.

\begin{table}[t]
	\renewcommand{\arraystretch}{1.5} 
	\begin{tabular}{c ll||c ll}
		\hline\hline
		& Benchmark Scenarios &&
		(e). & $(q_{e}^{},q_{n}^{},q_{p}^{})=(1,0,0)$, & (black solid) \\
		\hline
		(a). & $(q_{e}^{},q_{n}^{},q_{p}^{})=(0,0,1)$, & (green dashed) &
		(f). & $(q_{e}^{},q_{n}^{},q_{p}^{})=(1,0,1)$, & (green solid) \\
		\hline
		(b). & $(q_{e}^{},q_{n}^{},q_{p}^{})=(0,1,0)$, & (blue dashed) &
		(g). & $(q_{e}^{},q_{n}^{},q_{p}^{})=(1,1,0)$, & (blue solid) \\
		\hline
		(c). & $(q_{e}^{},q_{n}^{},q_{p}^{})=(0,1,1)$, & (red dashed) &
		(h). & $(q_{e}^{},q_{n}^{},q_{p}^{})=(1,1,1)$, &(red solid) \\
		\hline
		(d). & $(q_{e}^{},q_{n}^{},q_{p}^{})=(0,1,-1)$, &(purple dashed) &
		(i). & $(q_{e}^{},q_{n}^{},q_{p}^{})=(1,1,-1)$, &(purple solid)
		\\
		\hline\hline
	\end{tabular}
	\caption{Benchmark scenarios with different dark charge assignments of
		$(q_{e}^{},q_{n}^{},q_{p}^{})$ 
		as discussed in text, which correspond to the colored curves 
		(indicated in this Table) as shown the relevant figures.\
		The charges are expressed as 
		$(q_{e}^{},q_{n}^{},q_{p}^{})
		=(c_e^V,\hs 2c^V_u\!\!+\!c^V_d,\hs c^V_u\!\!+\!2c^V_d)$.} 
	\label{tab:benchmarks}
	\label{tab:1}
\end{table}
%

\vspace*{3mm}
\section{Electron Recoil Signals from Exothermic Inelastic Dark Matter}
\label{sec:ER-signals}
\label{sec:3}
\vspace*{1mm}

In this section, we proceed to discuss the electron recoil signals 
of exothermic inelastic DM scattering, 
in which a heavier DM component $X'$ de-excites to a lighter component $X$,
and deposits the energy from the mass-splitting into the kinetic energy 
of the final states.\ 
The electron recoil signatures may arise from direct DM-electron scattering 
or from the Migdal effect following the DM-nucleon scattering.

\subsection{Inelastic DM-Electron Scattering}
\label{sec:elec-scatter}
\label{sec:3.1}

The differential event rate of the DM-electron scattering with respect to 
the recoil energy $E_R^{}$ is given by\,\cite{Essig:2011nj}\cite{Bloch:2020uzh}:
\begin{eqnarray}
\frac{\d R_\text{ER}^{}}{\d E_R^{}}\,=\,
\frac{\,\rho_\text{DM}^{}\,}{m_X^{}}
\frac{\sigma_e^{}}{\,8m_e^2\,} \!\!\int\!\!\d^3 v \frac{\,f(\vec{v}\!+\!\vec{v}_e)\,}{v}
\!\sum_{n,\ell}\!\frac{1}{\,E_R^{}\!-\!E_{n\ell}^{}}
\!\int\!\! \d q\, q|f_{n\ell}(E_R,q)|^2,
\end{eqnarray}
where $\hs\rho_\text{DM}^{}\!\simeq\! 0.3\,\text{GeV}\hsm /\text{cm}^3$\, 
is the local DM density,
$\hs f(\vec{v})\hs$ the local DM velocity distribution function in the galactic frame, $\vec{v}_e$ the velocity of Earth,
$E_{n\ell}$ the energy level of atomic electron, 
and $|f_{n\ell}(E_R,q)|^2$ the atomic form factor.\
Here for simplicity we assume that the local DM consists of 
only the heavier DM component.

With the approximation that the out-going electron behaves like 
free plane wave\,\cite{Essig:2011nj}\cite{Bloch:2020uzh}, 
the atomic form factor can be estimated as
\begin{eqnarray}
|f_{n\ell}|^2 = \frac{\,2\ell\!+\!1\,}{2\hs\pi^3}
\frac{\,m_e^{}(E_R^{}\!-\!E_{n\ell}^{})\,}{q}
\!\!\int_{k_-}^{k_+}\!\!\d k\,k|\chi_{n\ell}(k)|^2 \,,
\end{eqnarray}
where the momenta $k_\pm^{}\!=\!|\sqrt{2m_e (E_R^{}\!-\!E_{n\ell})\,}\pm q|$ 
and $\chi_{n\ell}^{}$ denotes the radial wave function.
In this work, we adopt the numerical results for the radial wave functions given in\,\cite{wave-functions}.\ 
Beyond the plane wave approximation, the wave function of the ionized electron is affected by the attractive potential near the nucleus.\ 
This induces an additional enhancement to the above plane wave approximation,  
similar to the Sommerfeld enhancement.\ 
Following \cite{Essig:2011nj}, we multiply $|f_{n\ell}|^2$ by a Fermi factor which takes the following form in the non-relativistic limit,
\begin{eqnarray}
F_{\text{Fermi}}^{} \,=\,\frac{2\pi\eta}{~1\!-e^{-2\pi\eta}~}\,,
\end{eqnarray}
where 
$\hs\eta=Z_\text{eff}\hs\alpha\hs m_e^{}/\!
\sqrt{2m_e(E_R^{}\!-\!E_{n\ell}^{})\,}$.\ 
In this work we choose $\hs Z_\text{eff}^{}\!=\!1\hs$,  
which gives a relatively lower Fermi factor and 
thus predicts lower DM-electron event rates.\ 
Consequently, in this work the bounds on the DM-electron scattering 
are conservative bounds.

The DM velocity is assumed to follow a Maxwell-Boltzmann distribution in the galactic frame and truncated at the galactic escape velocity,  
$v_\text{esc}^{}\simeq0.00181$ \cite{McCabe:2010zh},
\begin{eqnarray}
	f(v) \propto
	\begin{cases}
		\exp(-v^2/v_0^2)-\exp(-v_\text{esc}^2/v_0^2), 
&~~ v\leqq v_\text{esc}^{}\,, 
\\[1mm]
0, &~~ v>v_\text{esc}^{}\,,
	\end{cases}
\end{eqnarray}
where $v_0^{}\!\simeq\! v_e^{}\!\simeq\! 0.00073\hs$ 
is the circular speed of the solar system around the galactic center.\
The minimum DM velocity for a scattering event to occur with electron recoil 
energy $E_R^{}$ and momentum transfer $q$ is given by 
$\,v_\text{min}\!=\!|q^2\!+\!2m_X^{}(E_R^{}\!-\!\Delta m)|/(2q\hs m_X^{})\hs$.\ 
Since the integral
\begin{equation}
	\xi(E_R,q) \equiv \int_{v_\text{min}}^{\infty} \!\!\d^3 v \frac{~f(\vec{v}+\hsm\vec{v}_e)~}{v}\,
\end{equation}
has an analytic form\,\cite{McCabe:2010zh,Savage:2006qr} 
as a function of the recoil energy $E_R^{}$ and momentum transfer $q\hs$, 
it is practically more convenient to evaluate the following integral 
for the recoil spectrum,
\begin{eqnarray}
\label{eq:DM-e}
\frac{\d R_\text{ER}^{}}{\d E_R^{}}\,=\,
\frac{\,\rho_\text{DM}^{}\,}{m_X^{}}\frac{\,\sigma_e^{}\,}{\,8m_e^2\,}\! \sum_{n,\ell}\!\frac{1}{\,E_R^{}\!-\!E_{n\ell}^{}\,}
\!\!\int\!\! \d q\,q\, \xi(E_R^{},q) |f_{n\ell}^{}(E_R^{},q)|^2\,.
\end{eqnarray}

For the present study of the inelastic DM-electron scattering, 
we have $m_X^{}\!\gg\! m_e^{}$ and $p_X^{}\!\gg\! p_e^{}\,$.\ 
Thus, according to \cite{He:2020wjs}
the range of momentum transfer lies in the range, 
$q_-^{} \!\leqq\hsm q \hsm\leqq\! q_+^{}$, with
\begin{eqnarray}
	\frac{q_\pm^{}}{\,m_X^{}\,}=\left|v_\text{DM}^{}\pm
	\sqrt{v_\text{DM}^2\!-2\!\left(\!\frac{\,E_R^{}\!-\!\Delta m\,}{m_X}
		\!\right)}\hs\right|,
\end{eqnarray}
where $v_\text{DM}^{}$ stands for the DM velocity.\ 
As shown in \cite{Bloch:2020uzh}\cite{Roberts:2016xfw}\cite{Roberts:2019chv}, 
the atomic form factors are peaked at 
$\hs q\!\sim\!O(10)\hs\text{keV}$ (depending on the recoil energy $E_R^{}$).\ 
Thus, most of the observed events satisfy 
$\,q_-^{}\!\!<\!O(10)\hs\text{keV}$,\footnote{
Because of the small momentum transfer in most of the observable events, 
we can ignore relativistic corrections 
that could become important only for 
$q\gtrsim 500$\,keV \cite{Bloch:2020uzh,Roberts:2019chv}.}
which constrain the width of the electron recoil signal:
\begin{eqnarray}\label{eq:q-limit}
	\left|E_R^{}\hsm -\!\Delta m\right| \hs \approx \hs v_\text{DM}^{} \hs q_- \!< 0.1\hs\text{keV}.
\end{eqnarray}
This means that the electron recoil spectrum is sharply peaked at
$E_R^{}\!=\!\Delta m\,$.\ 
Also, since $|f_{n\ell}|^2$ is peaked, 
the integral in Eq.\,\eqref{eq:DM-e} is almost independent of $q_\pm^{}$ 
and hence $m_X^{}$ so long as Eq.\,\eqref{eq:q-limit} is satisfied.\ 
Thus the event rate given by Eq.\,\eqref{eq:DM-e} is proportional to $1/m_X^{}\hs$.

\begin{figure*}[t]   
\centering
\hspace*{-4mm}
\includegraphics[height=6cm]{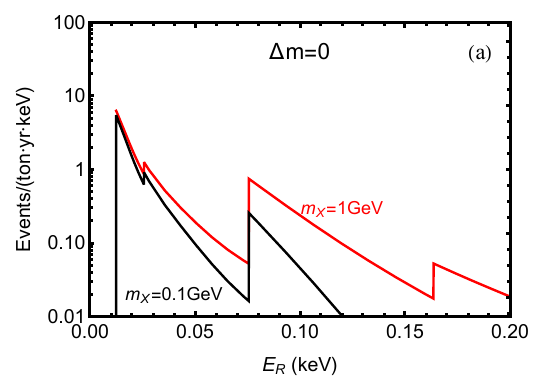}
\hspace*{-1mm}
\includegraphics[height=6cm]{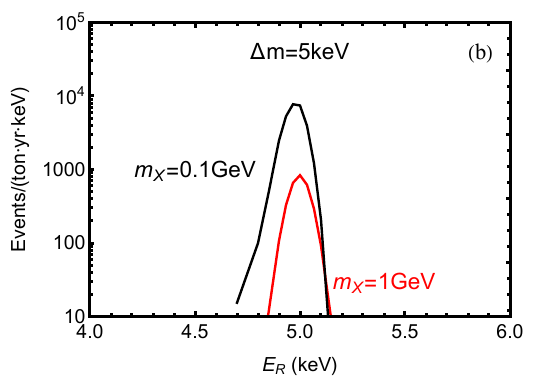}
\hspace*{-4mm}
\vspace*{-4.5mm}
\caption{\small
Electron recoil spectra by DM-electron scattering.\ 
The red and black curves show the ER spectra for DM mass 
$\mX=1\,\text{GeV}$ and $\mX =0.1\,\text{GeV}$, respectively.\ 
Plot-(a): The elastic scattering case.\ 
Plot-(b): The inelastic scattering case with a mass-splitting 
$\Delta m \!=\!5\,\text{keV}$.\ 
In this case the spectra are not only much stronger, 
but also are in a more detectable region.\ 
It shows that the sub-GeV DM produces a much stronger spectrum than the GeV-scale DM.\  
}
\label{fig:DM-e-compare}
\label{fig:1}
\end{figure*}

In Fig.\,1, we show a comparison of the electron recoil spectra 
between elastic and inelastic DM-electron scatterings.\ 
We see that for the elastic case in Fig.\,\ref{fig:1}(a), since the recoil energy comes entirely from 
the kinetic energy of the incoming DM particle, 
we have $E_R^{}\!<\!\frac{1}{2}m_X v_\text{DM}^2$.\ 
For sub-GeV DM, such recoil energy is always less than $1\,\text{keV}$, 
thus its signal is always below the detection threshold\,\cite{Essig:2011nj}.\ 
Moreover, the signal is weaker for lighter DM mass, 
as shown by the	black curve in Fig.\,\ref{fig:1}(a).\ 
Thus, even if the detection threshold is low enough for sub-keV signals, 
the sensitivity still decreases quickly when DM mass falls below GeV scale.\ 
But for the inelastic scattering case in Fig.\,\ref{fig:1}(b), 
the above features do not hold.\ 
Firstly, the recoil energy is always around the mass-splitting $\Delta m$, 
so the signal is detectable as long as the mass-splitting is above keV scale.\ 
Secondly, since the recoil energy mainly comes from the mass-splitting rather than 
the kinetic energy of the incoming DM particle, 
the signal does not diminish for lighter DM mass.\ 
Actually, the event rate is enhanced for lighter DM mass 
because the DM number density is inversely proportional to the DM mass,  
$n_X^{}\!\propto\! m_X^{-1}$.\ 
In summary, the inelastic DM-electron scattering is highly detectable 
as compared to the usual case of elastic scattering,  
especially in the sub-GeV range.

\vspace*{1.5mm}
\subsection{Inelastic DM-Nucleus Scattering}
\label{sec:nuc-scatter}
\label{sec:3.2}
\vspace*{1mm}

For the inelastic DM-nucleus scattering with 
$m_X^{}\!\ll\! m_N^{}\hs$, 
we can approximate $\hs q \hsm\sim\hsm m_X^{} v_\text{DM}^{}\hs$ and
\begin{eqnarray}
E_\text{NR}^{} \,\sim \frac{q^2}{\,m_N^{}\,} 
\!+\! \frac{\,m_X^{}\,}{\,m_N^{}\,}\Delta m \,.
\end{eqnarray}
Its contribution to the electron recoil is further suppressed by a quenching factor $\mathcal{L}$\,,
\begin{eqnarray}
\label{eq:E-NR}
E_{\text{NR}^{}\to\text{ER}}^{} \,=\,  \mathcal{L}\,E_\text{NR}^{}\,.
\end{eqnarray}
As will be discussed in Section\,\ref{sec:4}, 
for the present study we focus on the case with DM mass range $(0.1\!-\!10)\hs$GeV, 
thus the $E_{\text{NR}\to\text{ER}}$ contributions are always below the 
detection threshold, whereas the direct nuclear recoil signals 
are significant only when $\mX$ is large.

In a DM-nucleus scattering event, the recoiled nucleus acquires a small velocity relative to its surrounding electron cloud; 
this may lead to ionization of electrons in the outer orbits, 
as described by the Migdal effect\,\cite{Migdal}\cite{Ibe:2017yqa}.\
Exothermic scattering may enhance the Migdal effect\,\cite{He:2020sat}\cite{Bell:2021zkr}\cite{Li:2022acp}
by releasing the extra energy $\Delta m$ into the ionized atomic electrons,
\begin{eqnarray}
\frac{1}{2}\mu_N^{}v_{\rm DM}^2 + \Delta m 
\,=\, \frac{1}{2}\mu_N^{}v^{\prime\hs 2} + E_{\rm ER}^{}\,,
\end{eqnarray}
where $v_{\rm DM}^{}$ is the DM velocity in the laboratory frame,
$v'$ the relative velocity after scattering,
and $E_{\rm ER}^{}\!=\!E_{n\ell}^{}\!+\!E_e$ the total electronic energy release, including the atomic ionization energy $E_{n\ell}^{}$ and the kinetic energy of the ionized electron $E_e\hs$.\ 
The kinematics determines the nuclear recoil energy,
\begin{equation}
E_{\rm NR}^{} \,=\, 
\frac{\,\mu_N^2v_{\rm DM}^2\,}{m_N^{}}
\Bigg[1\!+\!\frac{\,\Delta m \!-\!E_{\rm ER}^{}\,}{\mu_N v_{\rm DM}^2} \!-\!\sqrt{1\!+\!\frac{\,2(\Delta m \!-\!E_{\rm ER})\,}
{\mu_N v_{\rm DM}^2}\,}\cos\hsm\theta\Bigg],
\end{equation}
where $\theta$ is the scattering angle in the center-of-mass frame.\
For a given $E_{\rm ER}^{} \!>\! \Delta m\hs$,
there is a minimal DM velocity imposed by kinematics,
\begin{eqnarray}
\label{eq:vDM-min}
v_{\rm DM}^{\rm min}\,=\sqrt{\frac{~2(E_{\rm ER}^{}\!-\hsm\Delta m)~}{\mu_N}}.
\end{eqnarray}
We see that the minimal DM velocity is reduced due to the energy release 
$\Delta m\hs$.\ The inelastic scattering is always kinematically 
allowed so long as $E_{\rm ER}^{}\!\leqq\!\Delta m$
since this condition removes the lower bound 
\eqref{eq:vDM-min} on $v_{\rm DM}^{}\hs$.\ 

The differential electron recoil rate of the Migdal effect per unit target mass 
with respect to $E_{\rm ER}^{}$ is related to 
the nuclear recoil rate ($R_{\rm NR}^{}$)$\hs$:
\begin{eqnarray}
\label{eq:MIGD-ER}
\frac{{\rm d}R_{\rm MIGD}^{}}{{\rm d}E_{\rm ER}^{}\,} \,\simeq
\int\!\! {\rm d}E_{\rm NR} {\rm d}v\!
\left[\!
\frac{{\rm d} R_{\rm NR}^{}}{\,{\rm d}E_{\rm NR}^{} {\rm d}v\,}
\sum_{n,\ell} \!\frac{1}{\,2\pi\,}
\frac{\,{\rm d}P^c_{q_e}\!(n,\ell\!\to\!E_{e})\,}{\,{\rm{d}}E_{\rm ER}^{}}
\!\right]\!,
\end{eqnarray}
where $P^c_{q_e}\!(n,\ell\!\to\!E_{e})$ is the probability of exciting an 
electron of level-$(n,\ell)$ to a free electron with kinetic energy 
$E_e$ \cite{Ibe:2017yqa}.\ 
The momentum transfer to the excited Migdal electron is 
$\,q_e^{} \!=\hsm 2m_e^2 E_{\rm NR}^{}/m_N^{}$.
Note that here $1/q_e^{}$ is much larger than the typical orbital radius 
of the atomic electron, and thus $\hs P^c_{q_e}\!\!\propto\!\! q_e^2\hs$ 
\cite{Ibe:2017yqa}.\ 
Hence, the enhanced $\hs q_e^{}$ due to the exothermic scattering with DM 
enhances the excitation probability.\ 
The differential nuclear recoil rate per unit target mass is given by
\begin{eqnarray}
\frac{{\rm d}R_{\rm NR}^{}}{{\rm d}E_{\rm NR}^{}{\rm d}v} \,\simeq\,  \frac{~\rho_\text{DM}^{}~}{\mX}
\frac{\sigma_\text{Xe}^{}}{~2\hs\mu_N^2~} 
\frac{\,f(v\!+\!v_e^{})\,}{v}\,.
\end{eqnarray}
The nuclear recoil also contributes to the detected electronic energy 
$\hs E_{\rm det}^{}\!\!=\!{\cal L}\hs E_{\rm NR}^{}\!+\! E_{\rm ER}^{}\hs$,
where ${\cal L}\approx 0.15$\, is the quenching factor parametrizing 
the transfer of nuclear recoil energy to the measurable electronic excitations.\ 
For the sub-GeV DM and an electron recoil detection threshold around $0.1$\,keV,
the contribution from the quenching of nuclear recoil would be negligible.\ 
Nevertheless, we include this effect in our numerical analysis.

\begin{figure*}[t]   
\centering
\hspace*{-4mm}
\includegraphics[height=6cm]{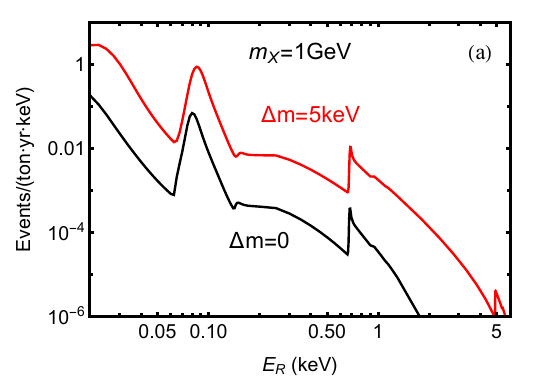}
\hspace*{-1mm}
\includegraphics[height=6cm]{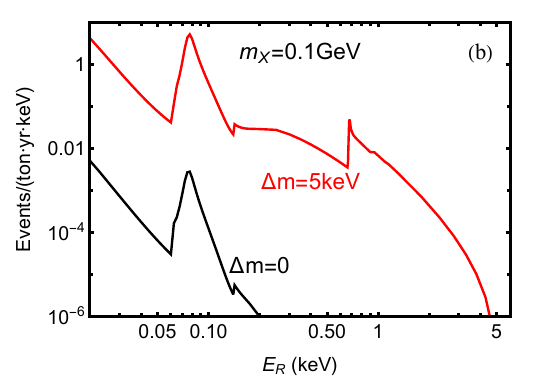}
\hspace*{-4mm}
\vspace*{-4.5mm}
\caption{\small
Electron recoil (ER) spectra from the Migdal effect by DM-nucleus scattering.\ 
The red and black curves present the ER spectra for elastic DM and inelastic DM with a mass-splitting $\Delta m =5\text{keV}$, respectively.\ 
Plot-(a) shows the case with DM mass $\mX\!=\!1$\,GeV, whereas 
plot-(b) depicts the case with DM mass $\mX\!=\!0.1$\,GeV. 
}
\label{fig:migd-compare}
\label{fig:2}
\end{figure*}

We compare the Migdal electron spectra from elastic and inelastic DM-nucleus scattering 
in Fig.\,\ref{fig:2}.\ 
The plots (a) and (b) present the electron recoil spectra for 
the cases of $\mX \!=\!1$\,GeV and $\mX \!=\!0.1$\,GeV respectively.\ 
The red curves correspond to the inelastic DM with $\Delta m\!=\!5\,\text{keV}$, 
whereas the black curves are for the elastic DM with $\Delta m\!=\!0\hs$.\ 
In each plot, we see that the signals are largely enhanced by the exothermic inelastic effect, especially in the region of $E_R^{}\!>\!1\,$keV.\ 
This is because the contribution from the mass-splitting $\delta m$ 
makes higher $E_\text{ER}^{}$ kinetically allowed.\ 
This guarantees that the signal never falls below the detection threshold for sub-GeV DM.\ 
For exothermic inelastic scattering, 
the event rate of the Migdal effect scales roughly as $\sqrt{1/m_X}$\,\cite{He:2020sat}.
In summary, we have demonstrated in Fig.\,\ref{fig:2} 
that the Migdal effect is enhanced by the exothermic effects of inelastic DM.

\vspace*{3mm}
\section{Limits from Direct Detection Experiments}
\label{sec:xenonnt}
\label{sec:4}
\vspace*{1mm}

In this section, using published data from the DM direct detection experiments, 
we analyze the electron recoil spectrum described in Section\,\ref{sec:ER-signals} and derive constraints on the inelastic DM.\ 
In the following analysis,
we consider the light inelastic DM in the mass range 
$(0.1\!-\!10)\hs$GeV.

\vspace*{2mm}
\subsection{Constraints by XENONnT Electron Recoil Data}
\vspace*{1mm}
\label{sec:4.1}

The first science run of XENONnT experiment provides 
the low-energy electron-recoil data\,\cite{XENON:2022ltv}
with an exposure of 1.16 tonne-years, which is currently the largest.\ 
It also has the lowest ER background rate 
in the $(1\!-\hsm 30)\hs$keV energy region.\ 
Thus, we use this dataset to constrain the effective cutoff scale $\Lambda$ 
of the inelastic interaction introduced in Section\,\ref{sec:DMmodels}.\ 
We extract the electron recoil data, background model $B_0^{}$, 
and detection efficiency $f_\text{eff}^{}(E)$ from Ref.\,\cite{XENON:2022ltv}.\
The detector smearing of the electron recoil spectrum is approximated 
by a skew-Gaussian distribution \cite{XENON:2022ivg},
\begin{eqnarray}
f(E_\text{d},E) \,=\, \frac{1}{\,\sqrt{2\pi\,} w\,} 
e^{-\frac{\,(E_\text{d}-\xi)^2}{2 w^2}}
\!\left[1\! + \text{erf}\left(\!\alpha 
\frac{\,E_\text{d}^{}\!-\!\xi\,}{\sqrt{2\,} w}\right)\!\right]\!,
\end{eqnarray}
where $E\hs$ is the physical energy, $E_d^{}$ the detected energy, and
\beqs 
\begin{align}
\alpha(E) &\,=\, 2.41E^{-0.30} \hs , 
\\
w(E) &\,=\, 0.374\sqrt{E}+0.005E \,, 
\\
\xi(E) &\,=\, 
E-\sqrt{\frac{2}{\pi}}\frac{\,\alpha w}{\sqrt{1\!+\!\alpha^2\,}\,}\,.
\end{align}
\eeqs 
The function $\alpha(E)$ should not be confused with the fine structure constant 
$\alpha\!=\!e^2/4\pi\,$ used in this paper.\ 
Thus, the event number detected per keV per ton-year is given by 
\begin{eqnarray}
\frac{\d N}{\d E_\text{d}^{}} \,=\, 
N_T^{}\!\!\int\!\!\frac{\d R}{\d E_R^{}}f(E_\text{d},E_R)
f_\text{eff}^{}(E_R)\d E_R^{}\,,
\end{eqnarray}
where $N_T^{}$ is the number of target xenon atoms per ton,
$E_R$ the (physical) recoil energy and $E_\text{d}$ the detected energy.\
The physical rate $\d R/dE_R^{}$ for DM-electron scattering and the Migdal effect
are given in Eqs.\eqref{eq:DM-e} and \eqref{eq:MIGD-ER}, respectively.\ 
The nuclear recoil contribution to the electron recoil data 
[as shown in Eq.\eqref{eq:E-NR}] is below the detection threshold, 
given that the DM mass $\,m_X^{}\!\leqq\!10\,$GeV and the quenching factor $\mathcal{L}\!=\!0.15\hs$. 
Thus it can be neglected in this analysis.

\begin{figure*}[t]   
\centering
\hspace*{-4mm}
\includegraphics[height=6cm]{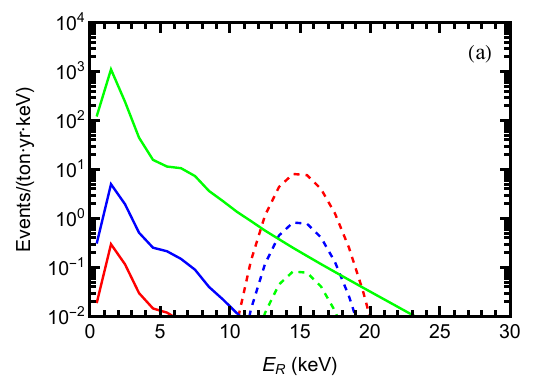}
\hspace*{-1mm}
\includegraphics[height=6cm]{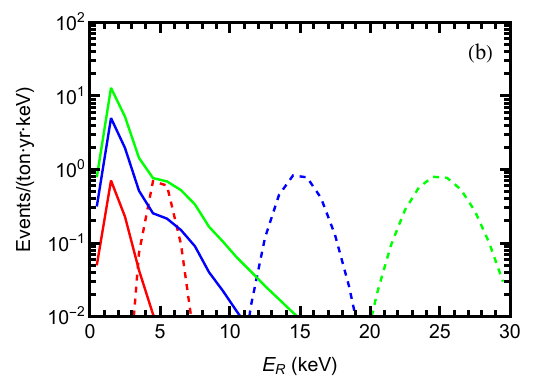}
\hspace*{-4mm}
\includegraphics[height=6cm]{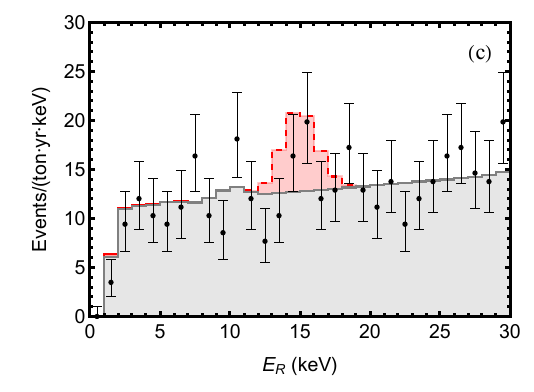}
\hspace*{-1mm}
\includegraphics[height=6cm]{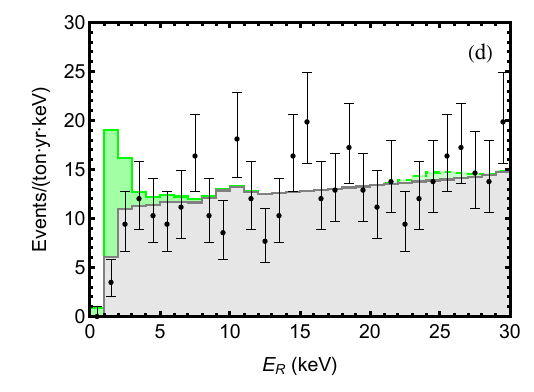}
\hspace*{-4mm}
\vspace*{-4.5mm}
\caption{\small
Electron recoil signals for XENONnT detection as contributed 
by the inelastic DM scattering.\
The XENONnT ER data\,\cite{XENON:2022ltv} are shown in plots\,(c)-(d)
as the black points (with error bars), 
and the backgrounds are marked by gray color.\ 
The solid curves arise from the Migdal effect, and the dashed curves 
are contributed by the DM-electron scattering.\ 
We set the cutoff scale $\Lambda \!=\hsm 350\hs$GeV
[cf.\ Eqs.\eqref{eq:sigma-N}-\eqref{eq:sigma-e}], 
and choose the dark charges $(q_e^{},q_n^{},q_p^{})\!=\!(1,1,-1)$.\
Plot-(a) presents the ER signal spectra with a sample mass-splitting 
$\Delta m\!=\!15\hs$keV, whereas
the (red,\,blue,\,green) curves present results for the DM mass 
$\hs \mX\!=\hsm (0.1,1,10)\hs$GeV, respectively.\
Plot-(b) presents the ER signal spectra with a sample DM mass  
$m_X^{}\!\!=\!\!1\hs$GeV, whereas 
the (red,\,blue,\,green) curves present results for 
$\Delta m=(5,\hs 15,\hs 25)\hs$keV, respectively.\
Plot-(c) presents the expected total ER spectrum in the detector 
with the same input as the red curves in plot-(a).\ 
Plot-(d) presents the expected total ER spectrum in the detector 
with the same input as the green curves in plot-(b).\ 
}
\label{fig:ER-spec}
\label{fig:3}
\end{figure*}

Since the dataset we use is in the $1\,\text{keV}\!\leqq\! E_R \!\leqq\! 30\,\text{keV}$ region, 
and the DM-electron scattering signal is always peaked at $E_R=\Delta m$, 
we focus on mass-splitting $\Delta m$ in the range $(1\hsm -\hsm 30)$\,keV in this analysis.\ 
In Fig.\,\ref{fig:3}, we present the predicted electron recoil spectra 
for XENONnT detector.\ 
(The color of each curve in this figure is not related to 
those for the benchmark scenarios defined 
in Table\,\ref{tab:1} of Section\,\ref{sec:2}.)
The signals from the DM-electron scattering are shown as dashed curves,
whereas the spectra for the Migdal effect are depicted in solid curves.\ 
The gray curves in plots\,(c)-(d)
show the contribution of backgrounds.\ 
For the DM-electron scattering, 
their event numbers are proportional to the local DM number density, 
so they scale as $1/m_X^{}$.\ 
This feature is demonstrated by the (red,\,blue,\,green)
dashed curves in plot-(a) which correspond to the DM mass
$\hs \mX\!=\hsm (0.1,1,10)\hs$GeV, respectively.\
We see that the DM-electron scattering spectra are always sharply peaked at $\Delta m\hs$, 
which is a unique feature of the inelastic DM.\  
Plot-(c) shows that such a striking peak signal 
could be consistent with the observed data of XENONnT in some regions 
such as the case of $\,\Delta m\!\sim\!15$\,keV, 
shown as the red shaded curve on top of the backgrounds.\ 
In this plot, the DM mass is small ($m_X\!=\!0.1\hs$GeV), 
so that the DM-electron scattering contribution dominates over that of the Migdal effect.\  

On the other hand, the event number generated by the Migdal effect can be enhanced by larger energy deposit, 
which is induced by larger DM mass $m_X^{}$ [cf.\ solid curves in plot-(a)] 
and/or larger mass-splitting $\Delta m\hs$ [cf.\ solid curves in plot-(b)].\ 
The shape and position of the spectrum are almost fixed, 
as shown in plots\,(a) and (b).\ 
We see that most of the Migdal events are in the region $E_R\!<\!3\,$keV.\ 
Since the backgrounds are already above the observed data points in this region, 
the existing XENONnT data can already place sensitive constraints on 
the Migdal events for relatively large $m_X^{}$ and/or $\Delta m\hs$, 
such as the case in plot-(d) where $m_X^{}\!=\!1\hs$GeV and $\Delta m\!=\!25\hs$keV.\ 
Thus, comparing plots\,(c) and (d)
shows that for larger DM mass and mass-splitting, the Migdal effect 
gives the leading contribution to the electron recoil events,
whereas for smaller DM mass and mass-splitting, the DM-electron scattering becomes dominant.\

\begin{figure*}
	\centering
	\hspace*{-4mm}
	\includegraphics[height=5.5cm]{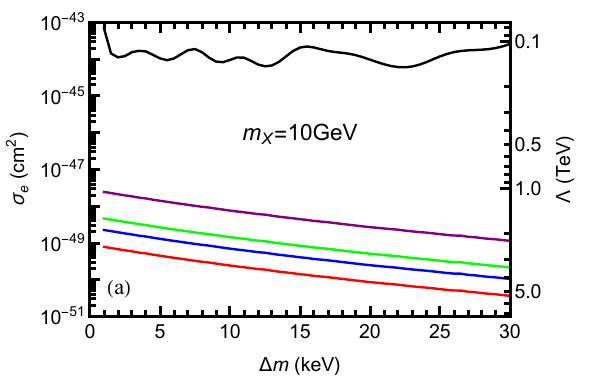}
	\includegraphics[height=5.5cm]{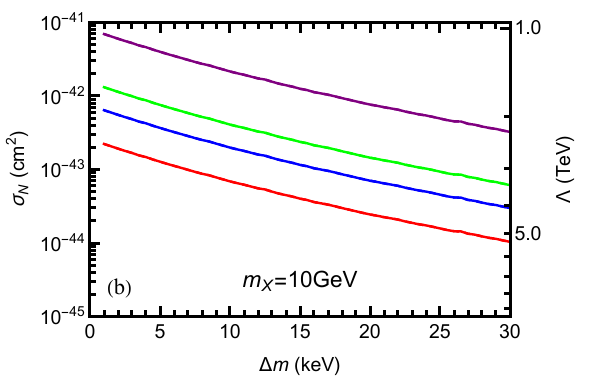}
	\hspace*{-4mm}
	\includegraphics[height=5.5cm]{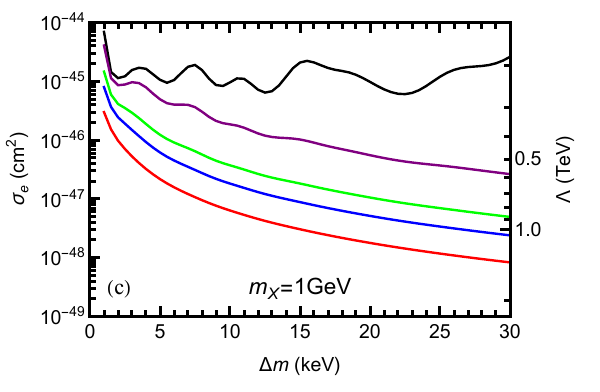}
	\includegraphics[height=5.5cm]{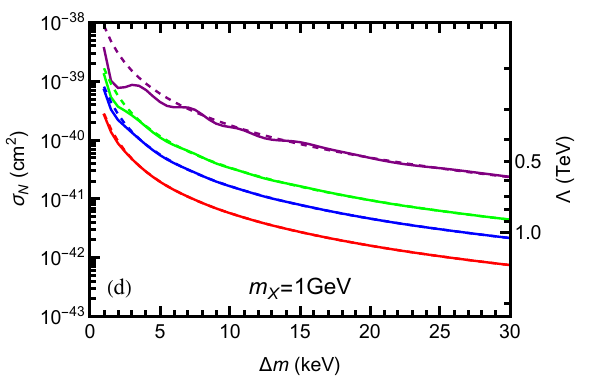}
	\hspace*{-4mm}
	\includegraphics[height=5.5cm]{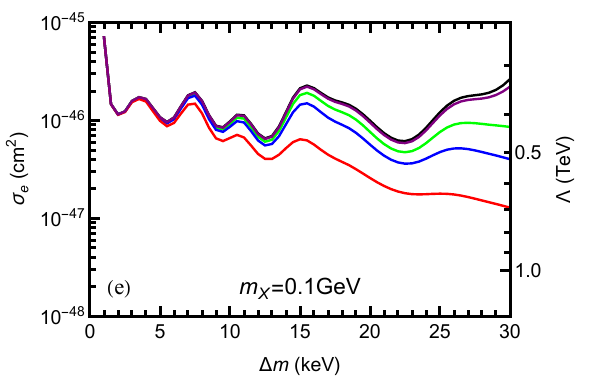}
	\includegraphics[height=5.5cm]{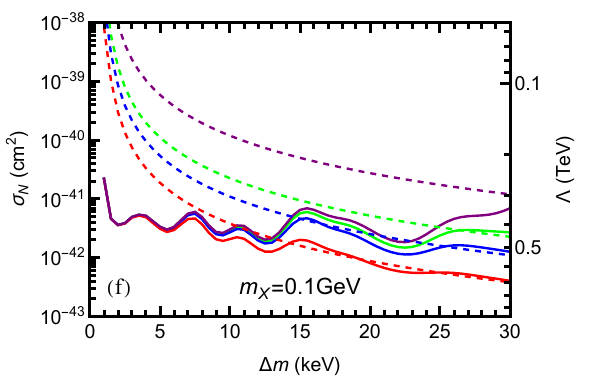}
	\hspace*{-2.5mm}
	\vspace*{-2.6mm}
\caption{\small
Bounds on the DM-electron and DM-nucleon scattering cross sections at 95\%\,C.L.\
from the XENONnT ER data, presented in the left and right panels respectively.\
The panels from top to bottom correspond to the DM masses 
$\mX \!=\!(10,\,1,\,0.1)\,${GeV}.\ 
The color of each curve matches the corresponding benchmark scenario (as indicated) in Table\,\ref{tab:1} (Section\,\ref{sec:2}),
i.e., dashed curves for $q_e^{}\!=\!0$ and solid curves for $q_e^{}\!=\!1$,
whereas green curves for $(q_n^{},q_p^{})\!=\!(0,1)$, blue curves for $(q_n^{},q_p^{})\!=\!(1,0)$, red curves for $(q_n^{},q_p^{})\!=\!(1,1)$, 
purple curves for $(q_n^{},q_p^{})\!=\!(1,-1)$, and black curves for 
$(q_n^{},q_p^{})\!=\!(0,0)$.\ The region above each curve is excluded.
}
\label{fig:ER-limits}
\label{fig:4}
\vspace*{3mm}
\end{figure*}

Throughout this analysis, we do not fluctuate the background,
because its normalization is fixed by the electron recoil data at 
$E_R^{}\!=\!(0\hsm -\!140)$\,keV, far beyond the range affected by DM scattering.\
We calculate the $\chi^2$ value for the background model $B_0^{}$ \cite{XENON:2022ltv} 
with $E_R^{}\!\leqq\!30$\,keV and obtain $\,\chi_\text{bkg}^2\!=\!19.84\hs$.\ 
Thus, at the 95\%\,C.L., we set 
$\hs\chi^2\!\leqq\! 19.84+3.84\hs$, which corresponds to one degree of freedom
(the free parameter $\Lambda\hs$).\ 
We plot the resulting limits in Fig.\,\ref{fig:4}.\ 
The (top,\,middle,\,bottom) rows correspond to the DM mass 
$m_X^{}\!=\!(10,\,1,\,0.1)\hs$GeV, respectively.\  
For each selected $m_X^{}$, we present the limits on the cross sections of 
the DM-electron (DM-nucleon) scattering in the left (right) panels,
and the corresponding limits on the cutoff scale $\Lambda\hs$.\ 
The color of each curve matches the corresponding benchmark scenario 
as defined in Table\,\ref{tab:1} of Section\,\ref{sec:2}.

\begin{figure*}[t]
	\centering
	\hspace*{-3mm}
	\includegraphics[height=5.8cm]{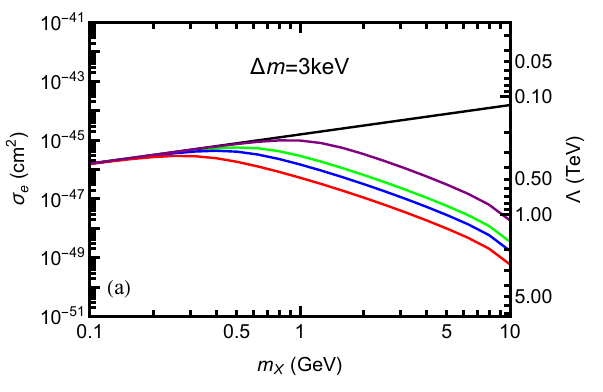}
	\includegraphics[height=5.8cm]{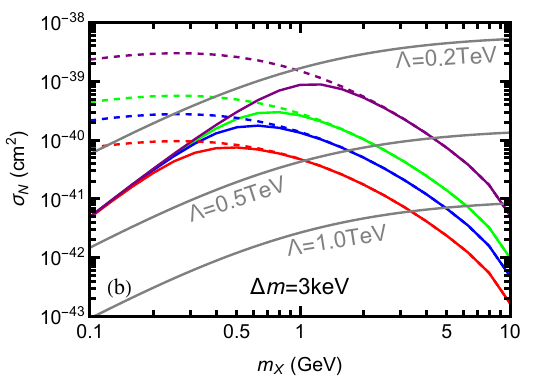}
	\hspace*{-3mm}
	\includegraphics[height=5.8cm]{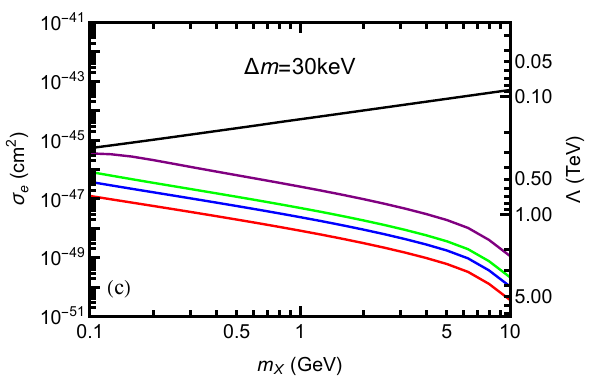}
	\includegraphics[height=5.8cm]{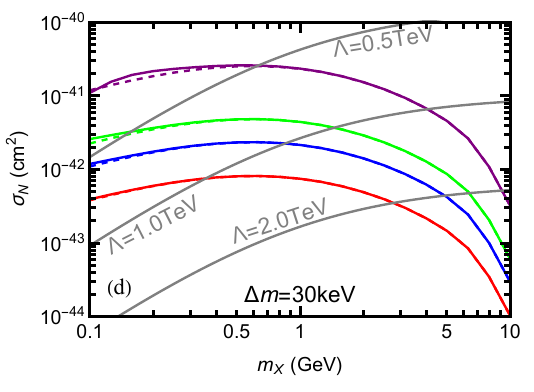}
	\hspace*{-3mm}
	\vspace*{-2.6mm}
\caption{\small
Bounds on the DM-electron and DM-nucleon scattering cross sections at 95\%\,C.L.\
by using the XENONnT ER data, presented in the left and right panels respectively.\
The upper (lower) panels correspond to $\Delta m\!=\!3\hs$keV ($30\hs$keV).
The color of each curve matches the corresponding benchmark scenario (as indicated) in Table\,\ref{tab:1} (Section\,\ref{sec:2}),
i.e., dashed curves for $q_e^{}\!=\!0$ and solid curves for $q_e^{}\!=\!1$,
whereas green curves for $(q_n^{},q_p^{})\!=\!(0,1)$, blue curves for $(q_n^{},q_p^{})\!=\!(1,0)$, red curves for $(q_n^{},q_p^{})\!=\!(1,1)$, 
purple curves for $(q_n^{},q_p^{})\!=\!(1,-1)$, and black curves for 
$(q_n^{},q_p^{})\!=\!(0,0)$.\ 
In plots\,(b) and (d), the gray curves from top to bottom correspond to 
$\Lambda\!=\!(0.2,\hs 0.5,\hs 1.0)\hs$TeV, respectively.\  
In all plots the region above each curve is excluded.}
\label{fig:ER-limits-new}
\label{fig:5} 
\vspace*{3mm}
\end{figure*}

The different impacts from the DM-electron scattering and from the Migdal effect 
can be seen by comparing the black curve (with $q_n^{}\!=\!q_p^{}\!=\!0$) 
to the colored curves (with $q_n^{}\!\neq\!0$ and/or $q_p^{}\!\neq\!0$) 
in the left panels of Fig.\,\ref{fig:4}.\ 
We note that some exclusion contours are smooth curves 
whereas others are twisty curves.\ 
The smoother ones correspond to the scenarios with the Migdal effect dominating 
the signal, i.e., when the DM mass is large (in the top panels except the 
black curve) and/or $q_e^{}\!=\!0$ (the dashed curves).\ 
This is because in such cases the shape of the total electron recoil spectrum 
is almost fixed, and the height of the peak is a smooth function of 
$\Delta m\hs$.\
Since most Migdal events are in the region of $E_R\!<\!3\hs$keV,
a low detection threshold for electron recoil is crucial to its detection.\ 
On the other hand, when DM-electron scattering is dominant 
(the solid curves in the bottom panels or the black curves),
the total electron recoil spectrum has peak-like structure 
whose peak positions depend on $\Delta m\hs$.\ 
Thus, the fluctuations among the data bins make each contour curve exhibit  
a distinctive twisty shape.\ A sufficiently high resolution 
in the electron recoil detection is required 
to determine the mass-spltting $\Delta m\hs$.\

We see that the Migdal effect generally dominates the recoil signals 
for $\mX \!\gtrsim\!1$\,GeV (top and middle panels of Fig.\,\ref{fig:ER-limits}) 
if the DM couples to nucleons.\ 
Even for smaller $\mX$, the Migdal effect can still be significant 
if the mass-splitting $\Delta m$ is large 
and increases the electron ionization probability 
through the enhanced energy release to electrons.\ 
This feature is evident in Fig.\,\ref{fig:5}.\
For large mass-splitting such as $\Delta m\!=\!30\,$keV 
in Fig.\,\ref{fig:5}(c)-(d),
the DM-electron scattering contribution is always subleading,
so that the colored curves ($q_n^{}\!\neq\!0$ and/or $q_p^{}\!\neq\!0$) 
always give much stronger constraints
than the black curve ($q_n^{}\!=\!q_p^{}\!=0$) in Fig.\,\ref{fig:5}(c),
and the differences between dashed curves ($q_e^{}\!\!=\!0$) 
and solid curves ($q_e^{}\!\!=\!1$) in Fig.\,\ref{fig:5}(d) is negligible.\ 
In Fig.\,\ref{fig:5}(a)-(b) with $\Delta m\!=\!3\hsx$keV,
the DM-electron scattering becomes dominant for $\mX\!\lesssim\! 1\hs$GeV.\ 
This makes the colored curves coincide with the black one 
in the region $\mX\!\lesssim\! 1\hs$GeV of Fig.\,\ref{fig:5}(a),
whereas large differences between the dashed and solid curves appear in the region
$\mX\!\lesssim\! 1\hs$GeV of Fig.\,\ref{fig:5}(b).\ 
Such observation signifies the necessity of analyzing the contributions of 
both DM-electron and DM-nucleus scattering simultaneously
because it is nontrivial to determine the relative impacts of the two contributions.

\vspace*{2mm}
\subsection{Constraints by S2-Only Data of XENON1T}
\vspace*{1mm}
\label{sec:4.2}

The S2-only data of XENON1T\,\cite{XENON:2019gfn} can also set non-trivial 
constraints on the DM models.\
In this study, we use the datasets, response matrices, 
and background models provided 
by XENON1T collaboration\,\cite{XENON:2019gfn}.\
According to \cite{XENON:2019gfn},
we will use the ER data and CEvNS background models 
for the present analysis, whereas the cathode backgrounds is ignored.\

\begin{figure*}[t]
	\centering
	\hspace*{-6mm}
	\includegraphics[height=6.3cm]{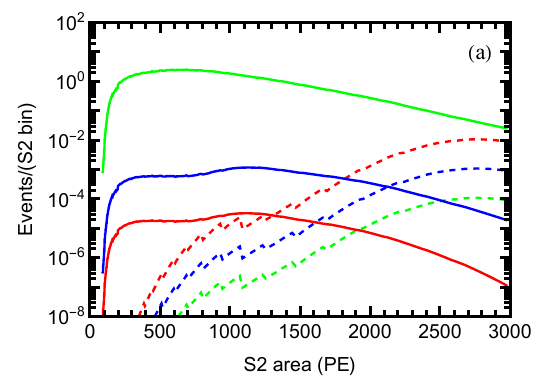} 
	\includegraphics[height=6.3cm]{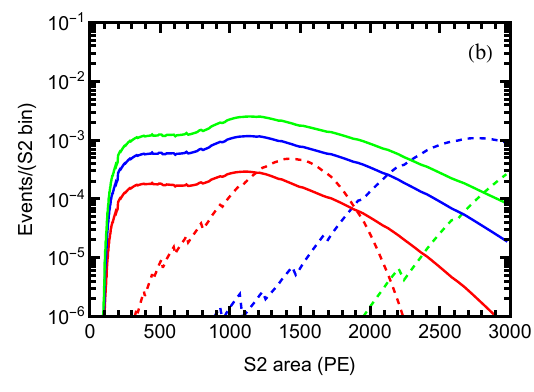}
	\hspace*{-4mm}
	\vspace*{-3mm}
\caption{\small 
Recoil signals from inelastic DM scattering by using the XENON1T S2-only data.\
The solid curves present contributions by the Migdal effect, and the dashed curves 
are given by contributions from the DM-electron scattering.\ 
We set the cutoff scale $\Lambda\!=\!500\hs$GeV, and choose the dark charges $(q_e^{},q_n^{},q_p^{})\!=\!(1,1,-1)$.\ 
Plot-(a): The mass-splitting is set as $\hs\Delta m\!=\!3$\,keV.\ 
The (red,\,blue,\,green) curves correspond to 
$\hs\mX\!\!=\!(0.1,\hs 1,\hs 10)\hs$GeV, respectively.\ 
Plot-(b): The DM mass is set as $\mX\!=\!1\hs$GeV.\ 
The (red,\,blue,\,green) curves correspond to  
$\Delta m\!=\!(1,\hs 3,\hs 5)\hs$keV, respectively.
}
\label{fig:S2only}
\label{fig:6}
\end{figure*}

The physical spectra of the electron recoil events include 
the DM-electron scattering and the Migdal effect, 
which are given by Eqs.\,\eqref{eq:DM-e} and \eqref{eq:MIGD-ER} respectively.\ 
To extract the S2 signals, we first apply the energy cutoff at $0.186$\,keV,
and then convert the physical event rate to the binned S2 rate 
using the response matrix.\ 
In Fig.\,\ref{fig:6}, we present a sample of the predicted S2 rates 
for the XENON1T detector with DM mass values,
$\mX\!=\!(0.1,\,1,\,10)\hs$GeV, respectively.\ 
Here the color of each curve does not correspond to the benchmark models (a)-(i)
defined in Table\,\ref{tab:1}.\ 
Note that the maximum S2 area is 3000 PE which translates to about 4\,keV,
thus for the current analysis of this subsection
the DM mass-splitting $\,\Delta m\,$ varies in the range 
$(0.1\hsm -\hsm 5)$\,keV.

The nuclear recoil contribution is always negligible for 
$\mX\!=\!0.1\hs$GeV and $\mX\!=\!1\,$GeV,
with a detection threshold of $O(\text{keV})$.\ 
As for $\mX\!=\!10\,$GeV, the nuclear recoil signal is significant
and the XENON1T collaboration has derived a bound, 
$\sigma_{N}^{}\!<\!4\times\! 10^{-45}\text{cm}^2$
for the elastic DM case\,\cite{XENON:2019gfn}.\ 
For the inelastic DM case with $\Delta m\!<\!5\hs$keV, 
we always have
\begin{equation}\label{eq:dm-to-NR}
\frac{\,\mX\,}{\,m_N^{}\,}\Delta m < {\rm keV},
~\quad \text{and}\quad~
\frac{\Delta m}{\,\mX v_\text{DM}^2\,} < 1\hs .
\end{equation}
This implies that the contribution from $\Delta m$ to nuclear recoil 
is always below the threshold
and sub-dominant as compared to the contribution from the kinetic energy 
of the incoming DM particle.\
Thus, the bound set by the XENON1T experiment\,\cite{XENON:2019gfn} should remain unchanged for the case of $\Delta m\!<\!5$\,keV.

\begin{figure*}
\vspace*{-8mm}
	\centering
	\hspace*{-4mm}
	\includegraphics[height=5.5cm]{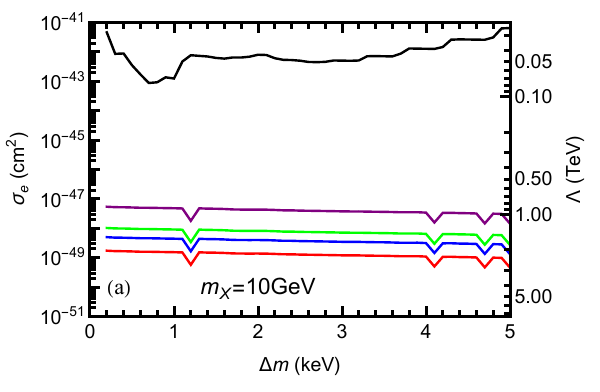}
	\includegraphics[height=5.5cm]{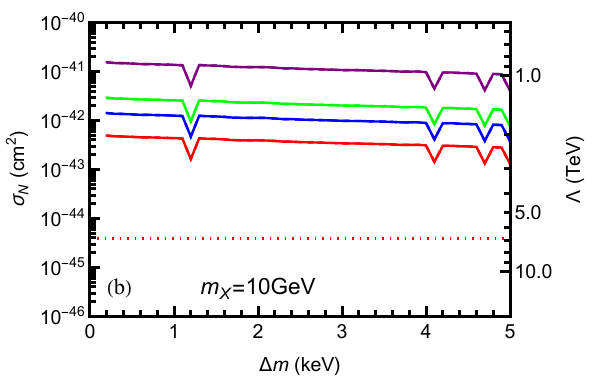}
	\hspace*{-4mm}
	\includegraphics[height=5.5cm]{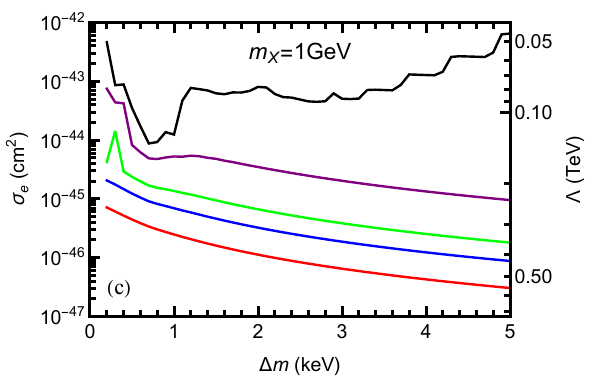}
	\includegraphics[height=5.5cm]{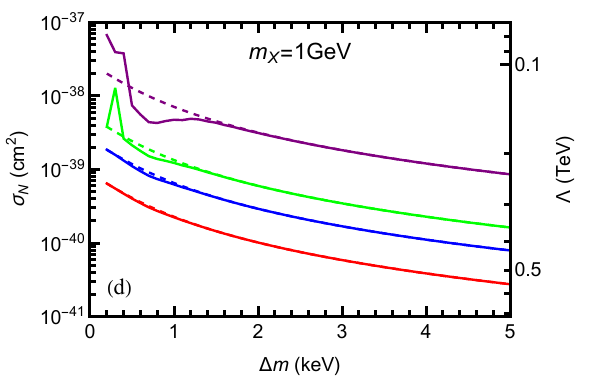}
	\hspace*{-4mm}
	\includegraphics[height=5.5cm]{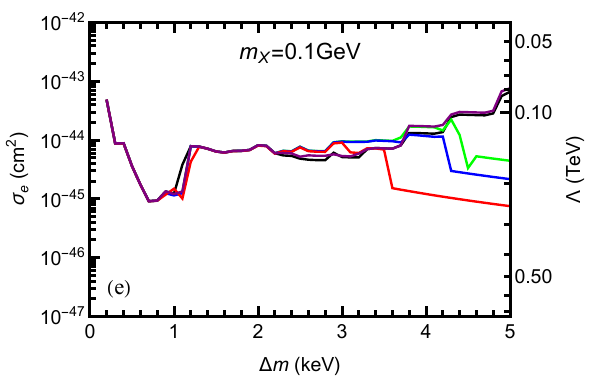}
	\includegraphics[height=5.5cm]{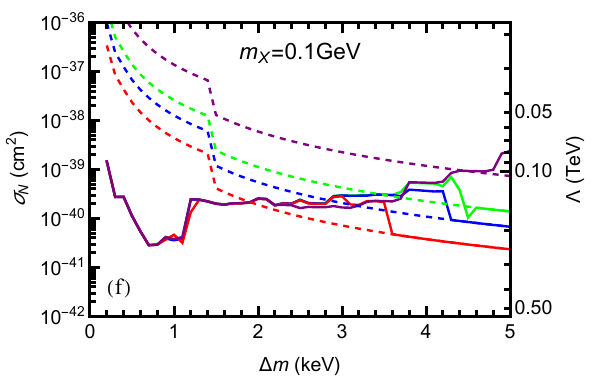}
	\hspace*{-2.5mm}
	\vspace*{-2.6mm}
\caption{\small
Bounds (95\%\,C.L.) on the DM-electron cross section (left panels) and 
on the DM-nucleon scattering cross sections (right panels)
by using XENON1T S2-only data.\ 
The plots from top to bottom choose the DM mass 
$\mX\!=\!\{10,\hs 1,\hs 0.1\}\hs$GeV, respectively.\ 
The color of each curve matches the corresponding benchmark scenario (as indicated) in Table\,\ref{tab:1} (Section\,\ref{sec:2}),
i.e., dashed curves for $q_e^{}\!=\!0$ and solid curves for $q_e^{}\!=\!1$,
whereas green curves for $(q_n^{},q_p^{})\!=\!(0,1)$, blue curves for $(q_n^{},q_p^{})\!=\!(1,0)$, red curves for $(q_n^{},q_p^{})\!=\!(1,1)$, 
purple curves for $(q_n^{},q_p^{})\!=\!(1,-1)$, and black curves for 
$(q_n^{},q_p^{})\!=\!(0,0)$.\
The region above each curve is excluded.
The red dotted curve in plot-(b) is taken from Ref.\,\cite{XENON:2019gfn}.\ 
It shows the 90\%\,C.L.\ upper limit on spin-independent DM-nucleon cross section 
by NR signals at $\mX\!=\!10\hs$GeV.\ 
}
\label{fig:S2-limits}
\label{fig:7}
\vspace*{3mm}
\end{figure*}

To derive the constraint for each model and 
each parameter set of $(\mX,\hs \Delta m)$,
we choose a custom S2 region of interest (ROI) 
before analyzing the search data of XENON1T.\  
As required in Ref.\,\cite{XENON:2019gfn}, the lower (upper) end of an ROI 
must lie between the 5th and 60th (40th and 95th) percentile of the S2 spectrum 
to contain sufficient signals, and should never be below 150\,PE.\
We first scan over all such ROIs on the training dataset and determine 
the optimized ROI that gives the lowest bound.\
Then, we apply this ROI on the search dataset\,(which is independent of the training dataset). The exclusion region (at 95\%\,C.L.) 
for the parameter $\boldsymbol{\theta}$ 
of a general probabilistic model is determined by the formula:
\begin{eqnarray}
	p(\boldsymbol{\theta}) \,\equiv\,  \sum_{n=0}^{n_\text{obs}}
	\frac{\,\mu(\boldsymbol{\theta})^n\,}{n!}
	e^{-\mu(\boldsymbol{\theta})}\leqq 0.05\,,
\end{eqnarray}
where $\hs\mu\,$ is the expected event number in the chosen ROI including 
the background and signal events,
and $\,n_{\text{obs}}^{}\hs$ denotes the observed event number in the chosen ROI.\ 
In the excluded region, the probability that more events would have been produced 
than what are actually observed (i.e., $n\!>n_\text{obs}^{}$) is above 95\%.\ 
For our scenario under consideration, there is only one free parameter 
$\hs\boldsymbol{\theta}=\Lambda$\,.

\begin{figure*}
	\centering
	\hspace*{-3mm}
	\includegraphics[height=5.8cm]{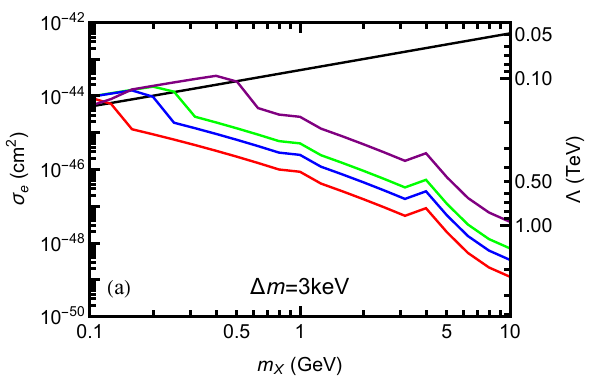}
	\includegraphics[height=5.8cm]{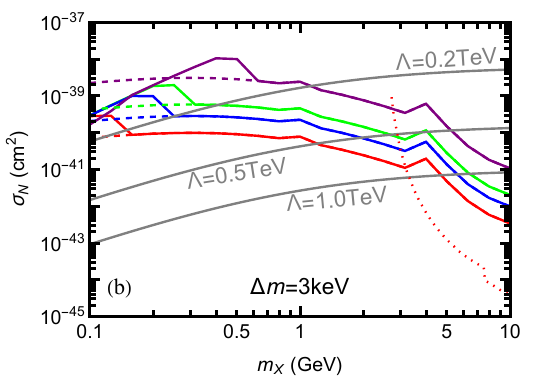}
	\hspace*{-3mm}
	\vspace*{-2.6mm}
	\caption{\small 
		Bounds (95\%\,C.L.) on the DM-electron cross section (left panel) and 
		on the DM-nucleon scattering cross sections (right panel)
		by using XENON1T S2-only data.\ 
		The color of each curve matches the corresponding benchmark scenario 
		(as indicated) in Table\,\ref{tab:1} (Section\,\ref{sec:2}),
		i.e., dashed curves for $q_e^{}\!=\!0$ and solid curves for $q_e^{}\!=\!1$,
		whereas green curves for $(q_n^{},q_p^{})\!=\!(0,1)$, blue curves for $(q_n^{},q_p^{})\!=\!(1,0)$, red curves for $(q_n^{},q_p^{})\!=\!(1,1)$, 
		purple curves for $(q_n^{},q_p^{})\!=\!(1,-1)$, and black curves for 
		$(q_n^{},q_p^{})\!=\!(0,0)$.\
		In plot-(b), the gray curves from top to bottom correspond to 
		$\Lambda\!=\!(0.2,\hs 0.5,\hs 1.0)\hs$TeV, respectively.\  
		The red dotted curve is taken from Ref.\,\cite{XENON:2019gfn}.\ 
		It stands for the 90\%\,C.L.\ upper limit on spin-independent DM-nucleon 
		cross section given by NR signals.\ 
		In both plots the region above each curve is excluded.
	}
	\label{fig:S2-limits-new}
	\label{fig:8}
	\vspace*{3mm}
\end{figure*}

In Fig.\,\ref{fig:S2-limits}, we present the results from analyzing the XENON1T 
S2-only data with inputs of $\mX\!=\!(0.1,\hs 1,\hs 10)\hs$GeV, respectively.
The plots are arranged in the same way as in Fig.\,\ref{fig:4}.
In addition, we derive the constraints for $\Delta m\!=\!3\hs$keV
over the DM mass range of 
$\hs 0.1\hs\text{GeV}\hsm \leqq\hsm \mX\hsm \leqq\! 10\hs$GeV,
as shown in Fig.\,\ref{fig:S2-limits-new}.\ 
We see that the same pattern appears as in the results 
from the XENONnT ER data, namely, 
the dashed curves are smooth, whereas for small DM mass $\mX$ 
the solid curves are twisty.\ 
Because the S2-only event numbers of XENON1T 
are much smaller than that in the ER data of XENONnT, 
the statistic fluctuations are more significant.\ 
This also leads to the non-smoothness of the ROI choices,
which further enlarges the fluctuations in these curves.\footnote{%
Most notably in Fig.\,\ref{fig:S2-limits}(f), 
for $2\hs\text{keV}\!\lesssim\!\Delta m\lesssim 4\hs\text{keV}$,
the ROIs for the solid curves (chosen by the training dataset) are in regions 
with higher S2 area.\ The search dataset in this region contains slightly more observed events above the backgrounds and imposes weaker limits than the 
dashed curves, whose ROIs are in the regions with lower S2 area 
which contains zero events in the search dataset\,\cite{XENON:2019gfn}.}\ 
As we have confirmed, the Migdal effect is more important than DM-electron 
scattering for larger $\mX$ and/or larger $\Delta m\hs$.\ 
The red dotted curves in Fig.\,\ref{fig:7}(b) and Fig.\,\ref{fig:8}(b) 
are the 90\%\,C.L.\ upper limit on spin-independent DM-nucleon cross section 
taken from Ref.\,\cite{XENON:2019gfn}.\ 
We use red color for the dotted curves 
because this limit corresponds to the $\hs q_n^{}\!=\!q_p^{}\!=\!1\hs$ case.\ 
This limit is determined solely by elastic nuclear recoil signals, 
which is neglected in this study.\ 
As discussed in Eq.\,\eqref{eq:dm-to-NR}, 
the limits are unchanged for inelastic DM with $\Delta m\!<\!5\,\text{keV}$.\ 
In Fig.\,\ref{fig:8}(b) we see that this limit gets weakened rapidly 
as $\mX$ decreases.\ 
In consequence, for larger DM mass ($\mX\!\gtrsim\! 5\hs\text{GeV}$), 
the nuclear recoil signals dominates over that of the electron recoils, 
whereas for $\mX\!\lesssim\! 5\hs\text{GeV}$, 
the nuclear recoil effects are negligible.

\vspace*{3mm}
\section{Probing Inelastic DM with Dark Photon Mediator}
\label{sec:models}
\label{sec:5}
\vspace*{1mm}

Based upon the above model-independent study,
we further explicitly construct a minimal inelastic DM model with dark photon mediator
via kinetic mixing, which can exhibit the main features as 
we studied in the previous sections.\
Then, we systematically derive the cosmological bounds and laboratory bounds
on this minimal inelastic DM model.

\vspace*{1.5mm}
\subsection{Model Realization of Inelastic DM and Observational Constraints} 
\label{sec:5.1}
\vspace*{1mm}

In this model, we extend the SM with a dark $U(1)_X^{}$ gauge group,
which connects to the SM through kinetic mixing between the dark $U(1)_X^{}$ 
gauge field $\bar{A}'_\mu$
and the SM $U(1)_Y^{}$ gauge field $B_\mu^{}\hs$\,\cite{Holdom:1985ag}.\
Thus, the kinematic terms of the dark $U(1)_X^{}$ and the SM $U(1)_Y^{}$ 
are given by
\beq
\label{eq:kin-mix-1}
\mathcal{L} ~\supset~ 
-\frac{1}{4}\bar{\mathcal{A}}'_{\mu\nu}\bar{\mathcal{A}}'^{\mu\nu}\! -\frac{1}{4}B_{\mu\nu}B^{\mu\nu}\! 
-\frac{\kappa}{2}\bar{\mathcal{A}}'_{\mu\nu}B^{\mu\nu} , 
\eeq
where the dark gauge field strength
$\hs\bar{\mathcal{A}}'_{\mu\nu}\!=\! 
 \partial_\mu^{}\bar{A}'_\nu -\partial_\nu^{}\bar{A}'_\mu\hs$.\

After the spontaneous breaking of the electroweak gauge group 
$SU(2)_L^{}\!\otimes\!U(1)_Y^{}$,
the neutral gauge boson component  
$W^3_\mu$ of $SU(2)_L^{}$ and the hypercharge gauge boson $B_\mu^{}$ of $U(1)_Y^{}$
form the mass eigenstates,
$\bar A_\mu\hsm \!=\! B_\mu\cos\theta_W^{}+W^3_\mu\sin\theta_W^{}$ 
and $\bar Z_\mu\!=\! W^3_\mu\cos\theta_W^{}\!-\!B_\mu\sin\theta_W^{}$,
where $\theta_W^{}$ is the weak mixing angle.\ 
Thus, including the $W^3_\mu$ part and all mass terms, 
we can rewrite the kinetic terms \eqref{eq:kin-mix-1} 
as follows:
\begin{align}
\label{eq:kinetix_mixing}
\mathcal{L}  ~\supset & 
-\frac{1}{4}\bar{\mathcal{A}}^{\prime}_{\mu\nu}\bar{\mathcal{A}}'^{\mu\nu}\!\! 
-\frac{1}{4}\bar{\cal A}_{\mu\nu}\bar{\cal A}^{\mu\nu}\!\! 
-\frac{1}{4}\bar{\cal Z}_{\mu\nu}\bar{\cal Z}^{\mu\nu}\!\! 
-\frac{\,\kappa\,}{2}\bar{\mathcal{A}}^{\prime}_{\mu\nu}
\big(\bar{\cal A}^{\mu\nu}\!\cos\hsm\theta_W^{}\! 
-\!\bar{\cal Z}^{\mu\nu}\!\sin\hsm\theta_W^{} \hsm\big) 
\nn \\
& + \frac{1}{2}m_{\bar{A}'}^2 \bar{A}'_\mu \bar{A}^{\prime\mu}
	+ \frac{1}{2}M_{\bar Z}^2 {\bar Z}_\mu {\bar Z}^\mu\,,
\end{align}
where the gauge field strengths
$\hs\bar{\mathcal{A}}_{\mu\nu}\!=\!
\partial_\mu^{}\bar{A}^{}_\nu\! -\partial_\nu^{}\bar{A}^{}_\mu\hs$
and
$\hs\bar{\mathcal{Z}}_{\mu\nu}\!=\!
\partial_\mu^{}\bar{Z}^{}_\nu\! -\partial_\nu^{}\bar{Z}^{}_\mu\hs$.\ 
Then, we can write the interaction terms of the gauge fields
$(\bar{A}'_\mu,\,\bar{A}_\mu^{},\,\bar{Z}_\mu^{})$ with matter currents:
\beq
\label{eq:Lint-A'AZ}
\mathcal{L}_\text{int}  ~\supset~  
g_X^{}\bar{A}'_\mu J_X^\mu + e\bar{A}_\mu^{} J_\text{em}^\mu 
+ \frac{g}{\,\cos\hsm\theta_W^{}\,} \bar{Z}_\mu^{} J_Z^\mu\,,
\eeq
where $J_X^\mu$ is the dark current 
as in Eq.\,\eqref{eq:dim6}, 
$J_{\rm em}^\mu$ the electromagnetic current, 
and $J_Z^\mu$ the weak neutral current.

To diagonalize these kinetic mixing terms and mass terms in Eq.\eqref{eq:kinetix_mixing}, 
we transform the gauge fields from the basis 
$(\bar{A}'_\mu,\,\bar{A}_\mu^{},\,\bar{Z}_\mu^{})$  
to the non-mixing mass-eigenbasis 
$(A'_\mu,\,A_\mu^{},\,Z_\mu^{})$.\ 
For the present study, 
we consider kinetic mixing parameter 
$\kappa\!\lesssim\! O(10^{-3})$ and 
dark photon mass $m_{\bar{A}'}^{}\!\lesssim\! O(\text{GeV})\hs$.\ 
Thus, to the leading order of the mixing parameter $\kappa$ and the mass ratio
$m_{\bar{A}'}^2/M_{\bar Z}^2$, we derive
\begin{eqnarray}
\label{eq:gauge_boson_mixing}
\begin{pmatrix}
\bar{A}'_\mu\hs \\[1mm]
\bar A_\mu^{}\hs \\[1mm]
\bar Z_\mu^{}\hs 
\end{pmatrix}
\simeq\, 
\(\hspace*{-2mm}
\ba{ccc}
1 & 0 & 
\kappa\hs s_W^{} 
\\[3mm]
-\kappa\hs c_W^{} & 1 & 0  
\\[2mm]
- \frac{m_{\bar{A}'}^2}{M_{\bar Z}^2}\hs\kappa\hs s_W^{} & 0 & 1
\ea\hspace*{-2mm}\)\!\!\! 
\begin{pmatrix}
\!A'_\mu \\
\!A_\mu^{} \\
\!Z_\mu^{}
\end{pmatrix}\!,
\end{eqnarray}
where we denote 
$(s_W^{},\hs c_W^{})\!=\!(\sin\theta_W^{},\hs \cos\theta_W^{})$
with $\theta_W^{}$ being the weak mixing angle.\
Accordingly, we derive the mass spectrum of the neutral gauge bosons 
$(A'_\mu,\,A_\mu^{},\,Z_\mu^{})$ as follows:
\beq 
\label{eq:gauge_boson_masses}
m_{A'}^{} \hs\simeq\hs   
m_{\bar{A}'}^{} \big(1\! +\hsm \kappa^2 c_W^2 \big)^{\!\frac{1}{2}} ,
~~~~~
m_A^{}\hs=0 \,,
~~~~~
{M}_Z^2 \hs\simeq\hs  
M_{\bar Z}^{} \big(1 \!+\hsm \kappa^2 s_W^2 \big)^{\!\frac{1}{2}} .
\eeq 
After diagonalization of the gauge boson mass-matrix, 
the interaction Lagrangian \eqref{eq:Lint-A'AZ} becomes:
\begin{align}
\mathcal{L}_\text{int}^{} ~\supset~ & 
g_X^{}\!\left(\!A'_\mu\!+ \kappa\hs s_W^{}\hsm  Z_\mu\hsm\right)\! J_X^\mu 
+ e\(A_\mu^{}\! - \kappa\hs c_W^{} A'_\mu\)\!J_\text{em}^\mu 
\nonumber \\
& + \frac{g}{\,c_W^{}\,}\!\hsm\(\hsm\!Z_\mu^{} \!-\! 
\frac{\,m_{\bar{A}'}^2\,}{M_{\bar Z}^2}\kappa\hs s_W^{}A'_\mu\!\)\!\! J_Z^\mu 
+ O(\kappa^2) \,.
\end{align}
In the following, we define $\hs\epsilon =\kappa\cos\theta_W^{}\hs$ 
as the effective mixing parameter.

For a complex scalar DM $\,\widehat{X}\!=\!(X\!+\!\ii X')/\!\sqrt{2}$\,
with a kinetic term $|D^\mu\! \widehat{X}|^2$, 
we easily deduce its $U(1)_X^{}$ gauge interaction terms:
\begin{eqnarray}
{\cal L}_{\text{int}}^{} & \supset &
g_X^{}
\big(X^\dagger
\partial^\mu X'\! -
{X'}^\dagger
\partial^\mu X\big)
\bar{A}_\mu^\prime \nonumber\\
& = & g_X^{}
\big(X^\dagger \partial^\mu X'\! - {X'}^\dagger \partial^\mu X\big)
\!\left(\!A'_\mu\!+ \kappa\hs s_W^{}\hsm  Z_\mu\hsm\right),
\label{eq:FDM_vertex}
\end{eqnarray}
which is consistent with the effective theory formulation in Section\,\ref{sec:2}.\ 
We note that the diagonal vertices
$X^{}$-$X^{}$-${A}^{\prime\mu}$ and
$X'$-$X'$-${A}^{\prime\mu}$ vanish,
whereas the above non-diagonal vertices can induce
the desired inelastic scattering.\ 
Similar behavior holds for the case of fermionic inelastic DM.

At leading order, the cross sections of DM-electron and DM-nucleon scattering  
are mediated by the dark photon $A'_\mu$.\ 
The contribution from $Z$-boson exchange is suppressed by 
$m_{A'}^4/m_{Z}^4\hs$.\ 
Thus, we derive the following cross section formulas:
\begin{subequations}
\begin{align}
\sigma_{XN}^{} & \,=\, Z^2 \frac{\,4\alpha g_X^2\epsilon^2\mu^2\,}{m_{A'}^4}\hs , 
\\
\sigma_{Xe}^{} & \,=\, \frac{\,4\alpha g_X^2\epsilon^2 m_e^2\,}{m_{A'}^4} \hs.
\end{align}
\end{subequations}
This should correspond to the scenario-(f) in Table\,\ref{tab:1} of Section\,2, 
where $q_n^{}\!=\!0$ and $|q_p^{}|\!=\!|q_e^{}|\!=\hsm 1\hs$,
and the effective cutoff scale is given by
\begin{eqnarray}
\label{eq:Lambda-mA'-gX-ep}
\Lambda\,=\,m_{\!A'}^{}/\!\sqrt{g_X^{}\hs \epsilon\,}\,.
\end{eqnarray}
In the above formula, we have considered that the momentum transfer 
$\hs q^2\!\ll\! m_{A'}^2\hs$.\ 
This is always true in the DM-electron scattering for 
$m_{A'}^{}\!>\! O(\text{MeV})\hs$, 
because the atomic form factors are peaked at 
$\,q\!\sim\! O(10\hs\text{keV})$ as we discussed earlier.\ 
But, for the DM-nucleon scattering, we have 
$\hs |q|\!\sim m_X^{} v_\text{DM}^{}\!\sim\! 10^{-3}\mX\hs$.\ 
Thus, we will focus on $m_{A'}^{}\!>\!10^{-2}\mX\hs$ in the following analysis.

Such exothermic inelastic scattering of DM particles with target atoms  
in the detector requires a significant local abundance 
of the heavier DM component $X'$.\ 
The conversion between $X'$ and $X$ is mainly due to $X'$ decay,
of which the only kinematically allowed channels are 
$\,X'\!\!\!\to\!\!X\gamma\gamma\gamma$ and $\,X'\!\!\to\!\!X\nu\bar\nu$.
Since the $3\gamma$ decay channel is one-loop suppressed,
the dominant decay channel is $\,X'\!\!\to\! X\nu\bar\nu\,$,
which occurs through the $A'/Z$ exchange due to the neutral gauge boson mixings 
in Eq.\eqref{eq:gauge_boson_mixing}.\ 
We can estimate the $X'$ decay width as follows:
\begin{align}
	\Gamma_{X'\rightarrow X\nu\bar\nu}^{} & \,=\,
	\frac{e^2 }{~1260\hs\pi^{3} M_Z^4 \cos^4\!\theta_W^{}\,} \frac{~g_X^2\hs\epsilon^2\,}{m_{A'}^4}\Delta m^9 \nonumber\\
	& \,\approx\,  \left(4\!\times\! 10^{34}\text{yrs}\right)^{\!-1}\!
	\(\!\!\frac{\Lambda}{\,100\hs\text{GeV}\,}\!\)^{\!\!-4}\!\!
	\(\!\frac{\Delta m}{\,10\hs\text{keV}\,}\!\)^{\!\!9} \,,
\end{align}
where we have included the contributions of the final state neutrinos 
from all three families of the SM.\
Note that this width is much narrower than that in our previous model\,\cite{He:2020sat},
because in this model the coupling between $A'$ and $\hs\nu\hs\bar{\nu}\hs$ 
is suppressed by $\hs m_{A'}^2/M_Z^2\hs$
and cancels the contribution from $Z$ exchange at leading order,
resulting in a tiny factor 
$\hs\Delta m^4\!/M_Z^4\hsm \simeq\hsm\!10^{-28}$  
in the decay width for $\Delta m\!=\!10\,$keV.\
This makes the lifetime of $X'$ much longer 
than the age of the universe.\ 
Hence, assuming equal initial abundance of $X$ and $X'$,
we have  $\rho_{X}^{}\!=\!\rho_{X'}^{}\!=\!\rho_\text{DM}^{}/2\,$ 
at the present.\ 
Consequently, the limits on the cutoff $\Lambda$ 
in Figs.\,\ref{fig:4}-\ref{fig:5} and Figs.\,\ref{fig:7}-\ref{fig:8}
should be rescaled by a factor of $\,2^{-1/4}\!\simeq 0.84\hs$.

This minimal inelastic DM model is also constrained by cosmological observations.\
The Planck Collaboration measured the 
present DM relic abundance\,\cite{Planck:2018vyg}, 
$\,\Omega_X^{} h^2 \!= 0.120\pm 0.001\hs$, 
which can be achieved through the freeze-out mechanism.\ 
Since $\Delta m\!\ll\! \mX$, the inelastic DM components 
$X$ and $X'$ are equally abundant in the thermal bath 
before exiting the equilibrium.\ 
For $\mX\!>\!m_{A'}^{}$, the dominant annihilation channel is $\hat{X}^\dagger\hat{X}\!\to\! A'A'$.\ 
This is similar to the secluded dark matter scenario\,\cite{Pospelov:2007mp}, 
where the parameter space for freeze-out is in general not limited by direct detection experiments.\ 
To be specific, the annihilation cross section scales as $\,g_X^4\hs$ 
and is independent of mixing parameter $\epsilon\hs$,
hence the bounds on $\Lambda\!\propto\!(g_X^{}\hs\epsilon )^{-1/2}$ 
from direct detection experiments
are irrelevant to the annihilation process.\
In fact, we can in turn derive the bounds on $\hs\epsilon\hs$ 
by combining the $g_X^{}$ values required by the DM relic density 
and the bounds on $\Lambda$ from direct detection experiments.\
But for $\mX\!<\!m_{A'}^{}$, the leading annihilation channel is $\hat{X}^\dagger\hat{X}\!\to\! A'^*\!\hsm\to\! f\hsx\bar{f}\hs$,
where $f$ and $\bar{f}$ are charged SM fermions.\  
The relic abundance of the DM can be estimated as follows\,\cite{Srednicki:1988ce,Gondolo:1990dk,PDG}:
\begin{equation}
\Omega_X h^2 \,\simeq\,
	0.1\!\(\!\!\frac{~x_f^{}\,}{20}\!\)\!\!\hsm 
	\(\!\!\frac{~10^{-8}{\rm GeV}^{-2}~}{\sigma_X^{}}\!\)\!,
	\label{eq:relic_eq}
\end{equation}
where $\hs x_f^{}\!\equiv\! T_f^{}/\mX\hs$ with $T_f^{}$ the freeze-out temperature.\ 
The DM annihilation cross section 
$\hs\sigma_X^{}\!\sim\!\alpha\hs m_X^2\hsm /\Lambda^4$\, 
is highly constrained by direct detection experiments,
since the green solid curves in 
Figs.\,\ref{fig:ER-limits}-\ref{fig:ER-limits-new} and 
Figs.\,\ref{fig:S2-limits}-\ref{fig:S2-limits-new} have set 
$\Lambda\!\gtrsim\! O(10^2\hs\text{GeV})$.\ 
This always results in an overproduction of DM, 
so that the case of $\mX\!<\!m_{A'}^{}$ is disfavored.\ 

\begin{figure*}[t]
	\centering
	\includegraphics[height=8cm]{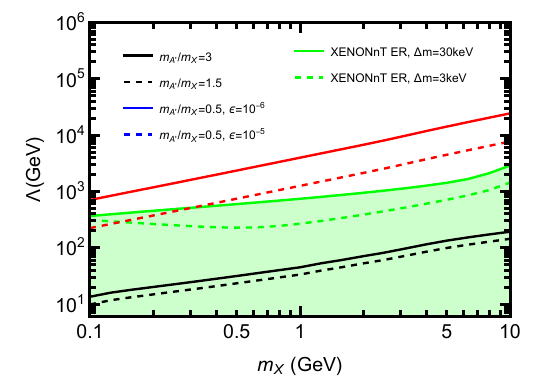} 
	\vspace*{-5mm}
	\caption{\small
		Allowed parameter space of $(\mX,\,\Lambda)$ as given by imposing  
		the DM relic density bound together with XENONnT electron recoil bounds of Fig.\,\ref{fig:ER-limits-new}.\
		The parameter space of the black and red curves can provide the observed 
		DM relic density today,
		whereas the green region displays the excluded parameter space at 95\%\,C.L.\
		The red dashed (solid) curve corresponds to the sample inputs of 
		$m_{A'}^{}/\mX \!=\!0.5$ and $\,\epsilon\hsm =\! 10^{-5}$ 
		($\hs\epsilon\hsm =\! 10^{-6}\hs$).\ 
		The black dashed (solid) curve corresponds to the sample input of
		$m_{A'}^{}/\mX \!=\!1.5$ ($\hs m_{A'}^{}/\mX \hsm\!=\!3\hs$);
		and the green dashed (solid) curve corresponds to the sample input of
		$\Delta m\!=\!3\hs$keV ($\hs\Delta m\!=\hsm 30\hs$keV$\hs$)
		for XENONnT ER bounds with the green shaded area being excluded.\  
	}
	\label{fig:relic-new}
	\label{fig:9}
	\vspace*{1mm}
\end{figure*}

To demonstrate this, 
we present in Fig.\,\ref{fig:9} the $\Lambda$ values required 
by the inelastic DM relic density 
and compare them to the constraints from direct detection experiments.\ 
The green dashed (solid) curve corresponds to the sample inputs of
$\Delta m\!=\!3\hs$keV ($\hs\Delta m\!=\!30\hs$keV$\hs$)
for XENONnT ER bounds with the green shaded area being excluded.\  
We compute the DM relic density numerically by using the package {\tt MicrOMEGAs} \cite{Belanger:2020gnr}.\ 
We see that for the case of $\hs\mX\!<\hsm m_{A'}^{}$, the sample input of 
$m_{A'}^{}/\mX \!=\!1.5$ ($\hs m_{A'}^{}/\mX \!=\!3\hs$)
is excluded by XENONnT bounds, as shown by the black dashed (solid) curve.\  
As mentioned earlier, the bounds on 
$\Lambda\!\propto\!(g_X^{}\epsilon)^{-1/2}$ 
are irrelevant to the case of $\hs\mX\!\!>\!m_{A'}^{}\hs$.\ 
In order to translate the values of $g_X^{}$ given by imposing DM relic density 
to $\Lambda$, we choose sample inputs of $\epsilon\!=\!10^{-5}$ and $10^{-6}$ 
which correspond to the red dashed and solid curves in Fig.\,\ref{fig:9},
respectively.\ 
Since $\epsilon$ does not receive any lower bound, 
we can always choose small enough values of $\hsx\epsilon\hsx$ such that
the red curve is fully above the green shaded region of Fig.\,\ref{fig:9}
and thus escape the constraints.\

\begin{figure*}[t]
\centering
\hspace*{-5mm}
\includegraphics[height=6.4cm]{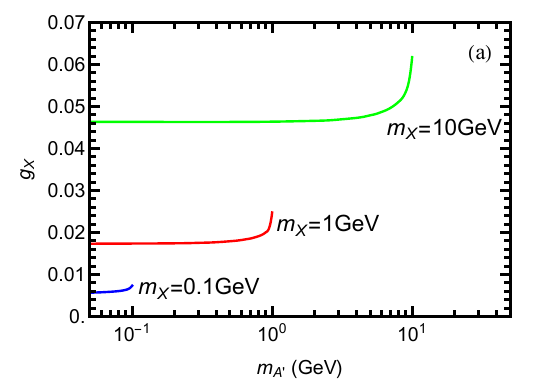}
\hspace*{-4mm}
\includegraphics[height=6.4cm]{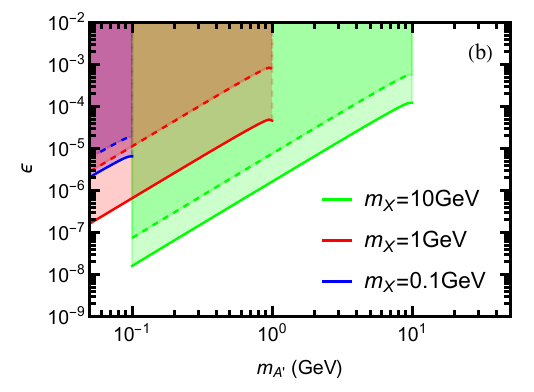}
\hspace*{-5mm}
\vspace*{-4mm}
\caption{\small
Plot-(a): 
Predicted $U(1)_X^{}$ coupling $g_X^{}$ versus the dark photon mass
$m_{A'}^{}$ as derived by imposing today's DM relic abundance 
$\Omega h^2\hsm\!=\hsm 0.120\hs$.\
Plot-(b):\ Bounds in the $(\epsilon,\,m_{\!A'}^{})$ plane.\ 
The (green,\,red,\,blue) colors denote the DM masses 
$\mX\!=\!(10,\,1,\,0.1)\hs$GeV, respectively.\ 
The solid (dashed) curves are derived from combining the DM direct detection 
and the DM relic density, with input $\Delta m\!=\!30\,$keV\,($1\hs$keV).\ 
}
\label{fig:relic}
\label{fig:10}
\vspace*{1mm}
\end{figure*}

In Fig.\,\ref{fig:10}(a), we present the $g_X^{}$ values that can give 
the correct DM relic density.\ 
Fig.\,\ref{fig:10}(b) shows the constraints on the dark photon 
from the direct detection experiments and cosmological observations.\ 
In this figure, the (green,\,red,\,blue) curves correspond to  
the DM mass $\hs \mX\!\!=\!(10,\, 1,\, 0.1)\hs$GeV, respectively.\ 
They are derived from $\hs\epsilon\!=\!g_X^{-1}m_{A'}^2\Lambda^{-2}$,
where the coupling $g_X^{}$ is constrained in Fig.\,\ref{fig:10}(a) 
as a function of $m_{A'}^{}$,
and the cutoff scale $\Lambda$ is constrained by the direct detection constraints 
as in Fig.\,\ref{fig:4}.\ 
The solid (dashed) curves corresponds to $\Delta m\!=\!30\,$keV\,(1$\hs$keV).\ 
Note that the dark photon mass $m_{A'}$ is always greater than 50\,MeV 
so that the $N_\text{eff}$ value during BBN is not affected.\ 
And the green curves in Fig.\,\ref{fig:10}(b) only cover the mass region 
$10^{-2}\mX\!\!<\!m_{A'}^{}\!\!<\!\mX\hs$, 
as is required in Section\,\ref{sec:5.1}. 

We see that for a given DM mass, the dark coupling $g_X^{}$ is a nonzero constant  
as $\,m_{\!A'}^{}\!\to\hsm 0\hs$.\ 
Thus, we derive the bound on $\epsilon$ [shown as solid and dashed curves in Fig.\,\ref{fig:10}(b)] which scales as $m_{\!A'}^2$ for small dark photon mass.\ 
This ensures a high sensitivity to the mixing parameter $\epsilon\hs$,
and allows a probe towards the parameter space 
$10^{-7}\!\!<\hsm\epsilon\hsm<\!10^{-3}$ and 
$0.01\hs\text{GeV}\!<\!m_{\!A'}^{}\!\!<\!10\hs\text{GeV}$,
where a large portion was previously not constrained by other experiments such as the fixed-target experiments
and the collider 
searches\,\cite{BaBar:2014zli}-\cite{LHCb:2019vmc}.\footnote{
A complete summary on the current bounds and the future projections on visibly decaying 
dark photon is given in Fig.\,128 of Ref.\,\cite{Antel:2023hkf}.}

Another cosmological constraint comes from the CMB anisotropy\,\cite{Slatyer:2009yq}\cite{Slatyer:2015jla}.\
After the DM freezes out, the annihilation is largely diluted 
but not completely shut down.\ 
The electron/positron injection from $A'$ decay 
contributes to the CMB anisotropy,
which is already constrained by the CMB measurements  
including Planck\,\cite{Planck:2018vyg}.\ 
The leading annihilation channel
$\hs\hat{X}^\dagger\hat{X}\!\to\! A'A'\!\to\!\ell\bar{\ell}\hs{\ell}\bar{\ell}\,$ 
is $s$-wave dominant.\ Its thermally averaged cross section receives 
an upper bound from the Planck data\,\cite{Planck:2018vyg}:
\begin{eqnarray}
	\label{eq:CMB_bound}
	f_{\rm eff}^{}\frac{\,\langle\sigma v\rangle\,}{\mX} 
	\,<\, 3.2\!\times\! 10^{-28}{\rm cm^3\hs s^{-1}GeV^{-1}},
\end{eqnarray}
where the efficiency factor $\,f_\text{eff}^{}\,$ 
is the fraction of the DM rest mass energy deposited into the gas\,\cite{Slatyer:2015jla}.\ 
In general, $f_\text{eff}^{}\!\gtrsim\! 0.1$ and it depends on 
the lepton energy injected into the gas.\ 
This excludes the full parameter space determined by the freeze-out analysis,
and would exclude the minimal model for the inelastic DM scenario with
dark photon mediator (via kinetic mixing).\
The same conclusion applies to the case of fermionic DM.\ 
But we note that the dark sector may contain several species of particles 
and mediators like the visible sector.\
Thus, in this case the CMB anisotropy constraint can be readily relaxed 
if $A'$ predominantly decays into light particles in the dark sector 
(rather than leptons)\,\cite{Bramante:2020zos}.\ 
We will give an explicit realization of this mechanism in the next subsection. 

Before concluding this subsection, we also comment on an inelastic DM model 
with dark photon mediator as introduced 
in Ref.\,\cite{He:2020sat}.\ 
In this model, the inelastic DM and the right-handed first generation SM fermions
are charged under a $U(1)_R$ gauge group.
The electroweak symmetry is spontaneously broken by two Higgs doublets, 
whose VEVs $(v_1^{},v_2^{})$ satisfy
$v_h^{}\!=\!\sqrt{v_1^2\!+\!v_2^2\,}\simeq\! 174\hs$GeV 
and $v_1^2\!\ll\! v_2^2\hs$.\ 
The dark photon mass is generated by the $U(1)_R^{}$ symmetry breaking.

The DM relic abundance can be provided by the freeze-out mechanism.\ 
To obtain the observed DM relic density,
a proper DM annihilation rate is needed.\ 
Since $\Delta m\ll m_X$, $X$ and $X'$ are equally abundant in the thermal bath 
before exiting the equilibrium.\
Thus, the dominant annihilation channel is 
$XX'\!\to\! A'A'\hs$ when $\hs m_{\!A'}^{}\! <\! \mX$
and $XX'\!\to\! A'\!\to\! f\bar{f}\hs$ when $\hs m_{\!A'}^{}\! >\! \mX\hs$.\ 
Since the interactions are suppressed by $\Lambda$, the annihilation rate is 
sufficient only when the channel 
$XX'\!\to\! A'A'$ is near the resonance during freeze-out.\ 
This requires $\hs m_{\!A'}^{}\!\gtrsim\hsm 2\hs\mX\hs$.\

After the freeze-out, the total DM abundance is already fixed,
but the ratio between $X$ and $X'$ can still evolve.\ 
As discussed in \cite{He:2020sat}, the leading channel for 
the conversion $X'\!\!\to\! X$ is
the decay $X'\!\to\! X \nu\bar{\nu}$ decay, which has the decay width
\begin{eqnarray}
\Gamma \,\simeq\, \frac{~q_\text{DM}^2g_X^4~}{160\hs\pi^3}
\frac{~v_1^4\Delta m_X^5~}{v_h^4m_{\!A'}^4}.
\end{eqnarray}
Ref.\,\cite{He:2020sat} sets the mass-splitting around 2.8\,keV,
whereas in the present study we allow the mass-splitting to vary within a much wider range
up to 30\,keV, which can bring the value of $v_1^{}$ down to $O(\text{GeV})$ scale
and has a higher VEV ratio $\tan\hsm\beta\!\lesssim\!\! 10^2$.\ 
A potential way to relax the decay width constraint is that 
the heavier state could be created by the terrestrial upscattering of the lighter state, 
so that the heavier state does not need to be abundant in the universe today\,\cite{Emken:2021vmf}.

Although the bounds in this work as imposed by current direct detection experiments 
are stronger than that of \cite{He:2020sat}, 
the collider constraints remain largely unchanged.\ 
As we discussed in \cite{He:2020sat}, 
the case of $m_{\!A'}^{}\hsm\!>\!2\mX$ is disfavored by collider searches.\ 

\vspace*{1.5mm}
\subsection{More on Model Realization}
\vspace*{1mm}

In this subsection, we give further constructions on this model  
for completeness of illustration,  
which are {\it non-essential} to realizing the main features of the inelastic DM as 
we studied in Sections\,\ref{sec:2}--\ref{sec:4} of this work.\  
These include to naturally generate the dark photon mass and 
the DM mass-splitting, as well as to provide the dark decay channel for dark photons.

The dark $U(1)_X^{}$ gauge group is spontaneously broken by 
the vacuum expectation values (VEVs) of two singlet scalar fields
$S$ and $\phi\,$,
\begin{align}
	\label{eq:L-scalar}
	\Delta\mathcal{L} ~\supset~ & 
	|D^\mu\! S|^2 \!+\! M_{S}^2|S|^2 \!-\!\lambda_S|S|^4  
	\!+\!|D^\mu\!\phi|^2 \!-\!M_{\phi}^2|\phi|^2 
	\!+\! \big(\lambda_{S\phi}S^3 \phi +\text{h.c.}\big)\hs ,
\end{align}
where the dark charges of $S$ and $\phi$ are $1/3$ and $-1$, respectively.\ 
This generates a mass term for ${\bar A'}_\mu$ through the Higgs mechanism, 
\begin{equation}
	\label{eq:DPmass}
	m_{\bar{A'}}^2 \,=\, 2\hs g_X^2\!\(v_\phi^2+\fr{1}{9}v_S^2\)\!,
\end{equation}
where {\,$g_X^{}\,$ is the $U(1)_X^{}$ coupling, }
and the VEVs are $\,v_S^{}\!=\!M_S^{}/\!\sqrt{\,2\hs\lambda_S^{}\hs}\,$ 
and $\,v_\phi^{}\!=\!\lambda_{S\phi}^{}v_S^3/M_\phi^2\,$.\ 
Taking $M_\phi^{}\!\gg\! v_S^{}\,$, 
we have $\hs v_\phi^{}\!\ll\! v_S^{}\hs$.\ 
This further gives $\,m_{\bar{A'}}^{}\!\simeq\!\sqrt{2}g_Xv_S/3$ 
and $v_\phi^{}\!=27\lambda_{S\phi}^{}m_{\bar{A'}}^3/(2\sqrt{2}g_X^3 M_\phi^2)\hs$.\ 

The inelastic DM consists of two nearly degenerate long-lived states.\
These states can be formulated by either a pseudo-complex scalar  
$\,\widehat{X}\!=\!(X\!+\!\ii X')/\!\sqrt{2}$\,
or a pseudo-Dirac spinor 
$\,\widehat{\chi}\!=\!(\chi_1^{},\,\chi_2^{\dag})^T$, 
which are charged under the dark gauge group $U(1)_X^{}$
with dark charge $+1$ for $\widehat{X}$ or 
dark charge $+\frac{1}{2}$  for 
$\widehat{\chi}^{}\hs$.\ 
(This means that the Weyl spinors $\chi_1^{}$ and $\chi_2^{}$ 
have dark charges $+\frac{1}{2}$ and $-\frac{1}{2}$ respectively.)
The mass-splitting $\Delta m\hs$ between the two DM components 
can be generated naturally by a seesaw mechanism.\ 
For the scalar dark matter $\widehat{X}$, the DM Lagrangian takes the
following form: 
%
\begin{align}
	\label{eq:L-DM}
	\Delta {\cal L}_\text{DM}^{} ~\supset~&
	|D^\mu\! \widehat{X}|^2 \!- m_{\widehat{X}}^2|\widehat{X}|^2\!
	-\lambda_X^{}|\widehat{X}|^4\!
	+\big(\lambda_{\delta}^{}\widehat{X}^2\!\phi^2
	\!+\! \text{h.c.}\big)
	\nn\\
	&
	- \lambda_{X\phi}^{}|\widehat{X}|^2|\phi|^2\!
	- \lambda_{XS}^{}|\widehat{X}|^2|S|^2\,.
\end{align}
The DM masses are mainly determined by the DM quadratic mass terms  
and the DM couplings to $|S|^2$ and $|\phi|^2$.\  
Thus, the DM mass terms are given by
\begin{align}
	\label{eq:Lmass-DM}
	\Delta {\cal L}_\text{DM} &~\supset~ 
	-\fr{1}{2}
	\big(m_{\widehat{X}}^2\!+\!\lambda_{XS}^{}v_S^2\!+\!\lambda_{X\phi}^{}v_\phi^2\big)
	\hsm 
	(X^2\!+\!X'^2)\hsm +\hsm\lambda_\delta^{}v_\phi^2\hs (X^2\!-\!X'^2)
	\nn\\
	&~=~  -\fr{1}{2} \over{m}_X^2(X^2\!+\!X'^2)
	+\fr{1}{2}\delta m^2 (X^2\!-\!X'^2)\,,
\end{align}
where the mass parameters 
$\hs\over{m}_X^{}\!=\!\!
(m_{\widehat{X}}^2\!+\!\lambda_{XS}^{}v_S^2\!+\!\lambda_{X\phi}^{}v_\phi^2)^{1/2}$ 
and
$\hs\delta m^2\!=\!2\lambda_\delta^{} v_\phi^2\!\ll\!\over{m}_X^2\,$.\ 
The mass-splitting $\Delta m_X^{}$
between the two DM components is determined by 
the unique quartic interaction $\widehat{X}^2\phi^2$
which gives rise to the mass-squared difference $\hs\delta m^2\hs$, 
as shown above.\ 
Since $v_\phi^{}\!\ll\! v_S^{},\over{m}_X^{}$ and thus 
$\hs\delta m^2\!\ll\!\over{m}_X^2\,$, we have
\\[-7mm]
\beq 
m_{X,X'}^{} \,\simeq\,
\over{m}_X^{}\mp \frac{\,\delta m^2\,}{~2\hs\over{m}_X^{}\,} \,. 
\eeq 
With these, we derive the scalar DM mass-splitting as follows:
%
\begin{align}
	\label{eq:mX'-mX}
	\Delta m_X^{} \simeq \frac{~\delta m^2\,}{~\over{m}_X^{}\,}
	\simeq
	\frac{~729\lambda_\delta^{}\lambda_{S\phi}^2 m_{A'}^6~}
	{4\hs g_X^6 M_\phi^4  \mX}\hs.
\end{align}
By setting the sample inputs 
$(\lambda_\delta,\hs\lambda_{S\phi}^{})\!=\!O(1)$, $g_X^{}\!=\!O(10^{-2})$, $(\mX,\,m_{A'}^{})=O(\text{GeV})$, and $M_\phi=O(10^2\hs\text{TeV})$,
we can achieve a naturally small mass-splitting 
$\Delta m_X^{}\!=\!O(\text{keV})\hs$.

For the fermionic DM $\widehat{\chi}$, 
the scalar potential \eqref{eq:L-scalar} remains the same 
and the fermionic DM has the following gauge-invariant Lagrangian terms:
\begin{align}
	\Delta{\cal L} \,\supset\,&~ 
	\chi^\dag_1\ii \bar\sigma^\mu \!D_\mu^{} \chi_1^{} +
	\chi_2^\dag\ii \bar\sigma^\mu \!D_\mu^{} \chi_2^{}
	-(m_{\widehat{\chi}}^{}\chi_1^{}\chi_2^{}\hsm +\text{h.c.})
	\nn\\
	&~
	+\big( y_{\phi\chi_1}^{} \chi_1^{}\chi_1^{}\phi
	+ y_{\phi\chi_2}^{} \chi_2^{}\chi_2^{}\phi^*\!
	+\text{h.c.} \big),
	\label{eq:FDM}
\end{align}
where the Majorana mass 
$m_{\widehat{\chi}}^{}$ and Yukawa couplings
$(y_{\phi\chi_1}^{}\hsm,\,y_{\phi\chi_2}^{}\!)$
are positive after proper phase rotations of
$\phi$ and $(\chi_1^{},\hs\chi_2^{})\,$.\ 
The small VEV $v_\phi^{}$ induces additional Majorana masses for
$(\chi_1^{},\,\chi_2^{})$
through the Yukawa interactions in Eq.\eqref{eq:FDM}.\
Thus, we have the following DM mass terms:
\begin{eqnarray}
	\mathcal{L}_{\chi_1\chi_2}^{} \,\supset\,  
	- m_{\widehat{\chi}}^{}\chi_1^{}\chi_2^{}
	\!+\hsm \delta m_1^{} \chi_1^{}\chi_1^{}
	+ \delta m_2^{} \chi_2^{}\chi_2^{} \!+\text{h.c.\hs},
\end{eqnarray}
where
$(\delta m_1,\,\delta m_2) \!=\!
(y_{\phi\chi_1}^{}\!\!v_\phi^{},\,y_{\phi\chi_2}^{}\!\!v_\phi^{})\hs$.\ 
To transform $(\chi_1^{},\hs\chi_2^{})$ into the mass-eigenstates
$(\chi^{},\hs\chi')$, we make the following decompositions:
\begin{eqnarray}
	\chi_1^{} = \frac{1}{\sqrt{2\,}\,}\big(\chi^{}\!-\!\ii\chi'\big),
	\hspace*{5mm}
	\chi_2^{} = \frac{1}{\sqrt{2\,}\,}\big(\chi\!+\!\ii\chi'\big).
\end{eqnarray}
Since $\,v_\phi^{}\!\ll\! m_{\widehat{\chi}}^{}$\,,
we derive the following Majorana masses for the
DM mass-eigenstates $(\chi,\hs\chi')\hs$:
\begin{subequations}
	\begin{align}
		m_{\chi}^{} &\,\simeq\, m_{\widehat{\chi}}^{} \!- ( \delta m_1\!+\!\delta m_2)\,,\\
		m_{\chi'}^{} &\,\simeq\, m_{\widehat{\chi}}^{} \!+ ( \delta m_1\!+\!\delta m_2)\,,
	\end{align}
\end{subequations}
which have a mass-splitting,
\begin{eqnarray}
	\Delta m_\chi^{} \simeq\,
	2( \delta m_1\!+\!\delta m_2)\,.
\end{eqnarray}
We choose the sample input parameters,
$\lambda_{S\phi}^{},\hs y_{\phi\chi_{1}^{}}^{}\!,\hs y_{\phi\chi_2^{}}^{}
\!\!=\! O(0.01)$,
$v_S^{}\!=\!O(20)\hs$GeV, and
$M^{}_\phi\!\!=\!O(\text{TeV})$.
With these inputs, we derive a small VEV $\,v_\phi^{}\!=O(0.1)$MeV,
and thus realize the desired mass-splitting
$\,\Delta m_\chi^{}\!=O(\text{keV})\hs$.

To relax the constraint from CMB anisotropies, 
we set that the dark photon decays dominantly into dark sector particles. 
For instance, the dark sector may contain a light Dirac fermion $f'$ 
charged under a new gauge group $U(1)_{X'}^{}$ with
gauge field ${A}''_\mu$\,.\ 
The gauge boson ${A}''_\mu$ obtains a mass $m_{{A}''}^{}$ 
through either the Higgs mechanism or the Stueckelberg mechanism, 
of which the detail is irrelevant to the discussion below.\ 
The kinetic mixing between $U(1)_{X'}^{}$ and $U(1)_Y^{}$ 
induces a small coupling between the electromagnetic current 
and ${A}''_\mu$ 
as parameterized by $\hs\epsilon' e \!\ll\! 1\hs$.\
The dark fermion $f'$ is charged under both $U(1)_X^{}$ and $U(1)_{X'}^{}$, 
with a mass obeying $m_{A'}^{}\!>\!2m_{f'}^{}$.\ 
The interactions of various currents takes the form,
\begin{equation}
{\cal L}\,\supset\, - g_{X}^{}\bar{f}'\gamma^\mu f' A'_\mu\!
- g_{X'}^{}\bar{f}'\gamma^\mu f' {A}''_\mu 
- \epsilon' e {\cal A}''_\mu J_{\rm em}^\mu\,,
\end{equation}
where $g_{X'}^{}$ denotes the gauge coupling of $U(1)_{X'}^{}$.\ 
The possible mixing between $U(1)_X^{}$ and $U(1)_{X'}^{}$ 
is set to be small (or vanishing), 
so the coupling between $\hat{X}$ and $A''$ is always 
suppressed and thus has negligible effect.\ 
In this setup, the ratio between the branching fractions 
Br[$A'\!\!\to\!\ell\bar{\ell}\hs$] and Br[$A'\!\to\! f'\bar{f}'$] 
is suppressed by the tiny factor of $\epsilon^2$, 
thus this setup can readily evade the bound 
from the CMB measurements.\footnote{
	The annihilation cross section of 
	$\hs XX'\!\!\to\! \ell\bar{\ell}\ell\bar{\ell}\,$ 
	through two virtual $A'$ is suppressed 
	by $\epsilon^4$ and satisfies the constraint \eqref{eq:CMB_bound} easily.}
Regarding the density of the stable particles $f'$, it is depleted 
by the efficient freeze-out process 
$\hs f'{\bar f}'\!\!\to\!\! A''\!\!\!\to\!\ell\hs\bar{\ell}\hs$.\ 
[Note that the annihilation $f'{\bar f}'\!\to\! \ell\hs\bar{\ell}\hs$ 
through a virtual $A'$ is insufficient to deplete the $f'$ density 
based on a similar relation to Eq.\eqref{eq:relic_eq}.]
This leads to $\hs n_{f'}^{}\!\ll\! n_X^{}$, 
so $f'$ contributes to neither the DM abundance nor the CMB anisotropy.\ 
Also, this setup brings a new annihilation channel 
$\hat{X}^\dagger\hat{X}\!\!\to\!A'^*\!\!\to\!\!f'\bar{f}'\hs$.\ 
For $\mX\!\!>\!m_{A'}^{}$, since it is $p$-wave dominant, 
i.e., suppressed by $\hs x_f^{}\!\sim\! 1/20\hs$ 
as compared to $\hat{X}^\dag\hat{X}\!\!\to\! A'A'$,
it barely affects the DM relic density.\ 
But for $\mX\!<\!m_{A'}^{}$, 
it becomes the dominant channel 
because it receives no suppression by $\epsilon^2$, as compared to 
$\hat{X}^\dagger\hat{X}\!\to\! A'^*\!\hsm\to\! f\hsx\bar{f}\hs$.\ 
Hence, unlike the minimal model, this setup allows the  
$\mX\!\!<\!m_{A'}^{}$ case to produce the correct DM relic density,
as well as avoiding the CMB anisotropy constraint.\ 
Hence, in this simple setup of the dark sector,
the interesting interplay between the constraints of the DM relic density 
and the direct detection still holds as in the minimal model,
whereas the cosmological bounds from CMB and BBN are satisfied.\ 

We re-compute the inelastic DM relic density with the extended dark sector.\ 
To be concrete, we choose the inputs $m_{f'}^{}\!=\!15\hs$MeV and $m_{A''}^{}\hsm\!=\hsm 45\hs$MeV, 
hence the freeze-out ends before the BBN at which $T\hsm\!\sim\hsm\! 1\hs$MeV.\ 
We set the coupling input $\hs g_{X'}^{}\!=\!\sqrt{2\pi\,}$ 
(and thus $\alpha_{X'}^{}\hsm\!=\! g_{X'}^2/4\pi\!=\!0.5\hs$)
and the mixing parameter $\epsilon'\!=\!10^{-4}$ 
(which is allowed by the bound of NA64 experiment\,\cite{NA64:2023wbi}).\ 
Throughout the parameter space, 
we affirm that the relic abundance of $f'$ is always small, $\hs\Omega_{f}^{} h^2 \!<\! 10^{-3}\hs$, 
hence it is consistent with the current observations.\ 
Also, it does not affect the limits presented in Fig.\,\ref{fig:10}. 
Note that the dark photon will decay invisibly, 
the bounds on visibly decaying dark photon\,\cite{BaBar:2014zli}-\cite{LHCb:2019vmc} 
do not apply here. 

\vspace*{1.5mm}
\section{Conclusions}
\label{sec:6}
\vspace*{1mm}

The origin and nature of dark matter (DM) remain largely unknown so far
except its participation in the gravitational interaction.\ 
It is therefore important to probe the DM particles for both large and small 
mass ranges, and through all possible means including the direct detection, 
indirect detection, and collider searches.

In this work, we focus on studying the light inelastic DM with mass around 
(sub-)GeV scale.\ 
In Section\,\ref{sec:2},  we gave a generic parameterization of 
the light inelastic DM interactions.\ 
For studying both the DM-electron scattering and DM-nucleus scattering 
(with Migdal effect) as well as their interplay, we presented the 
benchmark scenarios of the dark charge assignments 
as shown in Table\,\ref{tab:1}.\
Then, in Section\,\ref{sec:3}, we studied the electron recoil signatures 
from both the direct DM-electron scattering and the Migdal effect (induced by
DM-nucleus scattering), as shown in Fig.\,\ref{fig:1} and Fig.\,\ref{fig:2} respectively.\ 
We pointed out that unlike the case of the conventional elastic DM scattering, 
in which the recoil energy fully comes from the kinetic energy of the incoming DM particle, 
the exothermic inelastic DM provides extra energy from its mass-splitting 
to the final states.\ 
Thus, for the sub-GeV DM having too low kinetic energy, 
the inelastic effect could provide sufficient energy to trigger detectable signals.\ 
Since the local DM number density is proportional to $1/\mX$, 
the event rate for the sub-GeV DM is even larger than that of the GeV-scale DM.\  
Moreover, the inelastic DM-electron scattering predicts distinctive peak-like 
recoil spectrum that is highly detectable.

\vspace*{0.5mm}

In Section\,\ref{sec:4}, we analyzed the nature of the exothermic inelastic DM 
scattering process and demonstrated how direct detection experiments 
(such as XENON1T and XENONnT) can distinguish the signals from
the light inelastic DM, as shown in Fig.\,\ref{fig:3} and Fig.\,\ref{fig:6}.\ 
For the different benchmark scenarios of the dark charge assignments,
we derived the experimental bounds on the light inelastic DM 
from the XENONnT ER data and the XENON1T S2-only data
as shown in Figs.\,\ref{fig:4}-\ref{fig:5} and Figs.\,\ref{fig:7}-\ref{fig:8}.\ 
{\it These bounds are significantly stronger than 
those derived for conventional elastic DM,} 
because of the unique peak-like signal of DM-electron scattering
and the enhancement in the Migdal effect induced by DM-nucleus scattering.\ 
We found that these bounds depend sensitively on the dark charges 
$(q_e^{},\hs q_n^{},\hs q_p^{})$ and 
the mass parameters $(\mX,\hs \Delta m)$.\ 
For larger DM mass $\mX$ and/or vanishing dark charge $q_e^{}$, 
in which case the DM-nucleon scattering dominates,
the bounds are set mainly by the amplitudes of Migdal effect 
with the electron recoil energy around $(1\! -\hsmx 2)\hs$keV range,
whereas in the case with direct DM-electron scattering being dominant,
the signals are concentrated around the $E_R^{}\!=\!\Delta m\hs$ region.\ 
These facts suggest that a lower energy threshold in the detector 
can benefit the sensitivity for the DM-nucleon scattering with Migdal effect,
and a higher energy resolution can increase the sensitivity 
to the DM-electron scattering.\ 
The Migdal effect dominates over the direct electron scattering 
for larger $\mX$ or larger $\Delta m\hs$.\ 
But, for a general input of $(\mX,\hs \Delta m)$, we demonstrated that 
it is highly nontrivial to determine 
which effect is dominant.\ 
Hence, it is important to {\it perform a combined analysis of both effects.}

Finally, in Section\,\ref{sec:5},
we studied a model realization of such light inelastic DM with the
dark photon mediator, and applied the bounds of our benchmark scenarios 
to constrain this model as shown in Figs.\,\ref{fig:9} and \ref{fig:10}.\ 
We further studied the cosmological and laboratory 
constraints on the inelastic DM.\

Our analysis can be generalized to other direct detection experiments, 
especially LZ and PandaX-4T experiments, which also use liquid xenon as targets.\ 
It would be interesting for us to further study how their upcoming ER and S2-only data 
could be applied to probe the light inelastic DM.

\vspace*{5mm}
\noindent
{\bf\large Acknowledgements}
\\[1.5mm]
We thank Jingqiang Ye for discussing the XENON1T and XENONnT data analysis.\
This research was supported in part by the National NSF of China 
under grants 12435005, 12175136 and 11835005.


\baselineskip 17pt

\vspace{5mm}
%

\end{document}